\documentclass[11pt]{article}

\usepackage[utf8]{inputenc}
\usepackage{mathpazo}
\usepackage{aurical}
\usepackage{import}
\usepackage[toc,page]{appendix}\usepackage{latexsym,amsfonts,amsmath,
amssymb,mathrsfs,bbold,mathtools,esint,amsthm,mathtools}
\usepackage{dsfont}
\usepackage{mathtools}
\usepackage{float}
\usepackage{graphicx}
\usepackage{cite}
\usepackage{hyperref}
\usepackage{setspace}
\usepackage{color}
\usepackage{slashed}
\usepackage{cleveref}
\numberwithin{equation}{section}
\usepackage[colorinlistoftodos]{todonotes}
\usepackage[affil-it]{authblk}
\usepackage{float}
\usepackage{indentfirst}
\usepackage{soul}
\usepackage{booktabs}
\usepackage{eso-pic,graphicx}
\usepackage{nicefrac}
\usepackage{epsfig}

\newcommand{\normord}[1]{:\mathrel{\mkern2mu #1 \mkern2mu}:}

\newcommand{\delslash}{{\partial\mkern-9mu/}} %% usual Dirac operator
\usepackage[a4paper]{geometry}
\usepackage{a4wide}

\usepackage{color,hyperref}
\definecolor{darkblue}{rgb}{0.0,0.0,0.3}
\hypersetup{colorlinks,breaklinks,
            linkcolor=darkblue,urlcolor=darkblue,
            anchorcolor=darkblue,citecolor=darkblue}

\title{On the quantization of continuous non-ultralocal integrable systems}
\author[1]{A. Melikyan\footnote{\href{mailto:amelik@gmail.com}{amelik@gmail.com}}}
\author[2]{G. Weber\footnote{\href{mailto:gabrielweber@usp.br}{gabrielweber@usp.br}} }
\affil[1]{Instituto de Física\\
Universidade de Brasília\\
70910-900, Brasília, DF, Brasil}
\affil[2]{Escola de Engenharia de Lorena\\
Universidade de São Paulo\\
12602-810, Lorena, SP, Brasil}

\usepackage{palatino}

\begin{document}
\maketitle

\begin{abstract}
We discuss the quantization of non-ultralocal integrable models directly in the continuous case, 
using the example of the Alday-Arutyunov-Frolov model. We show that by treating fields as distributions 
and regularizing the operator product, it is possible to avoid all the singularities, and allow to obtain results
consistent with perturbative calculations. We illustrate these results by considering the reduction to the massive free fermion model and extracting the quantum Hamiltonian as well as other conserved charges directly from the regularized trace identities. Moreover, we show that our regularization recovers Maillet’s prescription in the classical limit.
%We discuss the quantization of non-ultralocal integrable models directly in the continuous case on example of the Alday-Arutyunov-Frolov model and its consistent reduction to the massive free fermion model. We show that by simultaneously regularizing the fields and the operator product, it is possible to avoid all the singularities in the diagonalization procedure and thus obtain results consistent with former perturbative calculations. We illustrate the usefulness of our proposal by extracting the quantum Hamiltonain as well as other conserved charges directly from the trace identities. Moreover, we show that our regularization recovers Maillet's prescription in the classical limit.

%We discuss the quantization of non-ultralocal integrable models directly in the continuous case, using the example of the Alday-Arutyunov-Frolov model. We show that by treating the fields as distributions one can obtain relations generalizing the Yang-Baxter relation for the ultralocal systems, which are free of singularities and allow to obtain results consistent with perturbative calculations. We also show that Maillet's symmetrization procedure is reproduced in the classical limit.
\end{abstract}

\paragraph{Keywords:} Exactly Solvable Models, Bethe Ansatz; Continuum models;
Integration of Completely integrable systems by inverse spectral and scattering methods; Quantum Field Theory.

\section{Introduction}

The Alday-Arutyunov-Frolov ($AAF$) model is a purely fermionic classical integrable model arising from the reduction of the $AdS_{5} \times S^{5}$ superstring theory to the $\mathfrak{su}(1|1)$ subsector in the uniform gauge \cite{Alday:2005jm,Arutyunov:2005hd}. Applying the inverse scattering method to the $AAF$ model has proven to be a very non-trivial task as a result of its non-ultralocality, which manifests in an even more complicated form than in the usual examples of non-ultralocal integrable systems \cite{Maillet:1985ek}. This prompts a more detailed investigation of non-ultralocal models and the development of new methods to regularize and therefore quantize such theories.

As a matter of fact, the quantization of non-ultralocal integrable models is one of the most intriguing and challenging open problems in the context of integrability. To date only in very few examples this question has been properly addressed, i.e., the $SU(2)$ Principal Chiral Model (PCM) \cite{Faddeev:1985qu}, the Wess-Zumino-Novikov-Witten (WZNW) model \cite{Alekseev:1992wn,Alekseev:1996jz} and the non-abelian Toda lattice field theory \cite{Freidel:1991jv}. Nonetheless, this problem has once more become a very active and relevant area of research since it was discovered that $AdS_{5} \times S^{5}$ string theory is a classically integrable system of this type (for a review see \cite{Beisert:2010jr} and references therein). In spite of all the attention devoted recently to this area, because of its importance to quantizing the $AdS_{5} \times S^{5}$ superstring and thus improving our understanding of the AdS/CFT correspondence \cite{Delduc:2012qb,Delduc:2012vq,Dorey:2006mx,Benichou:2010ts,Benichou:2011ch,Benichou:2012hc}, there is still no satisfactory general method to resolve all the difficulties involved in the quantization process of non-ultralocal theories.

There exist, though, some standard approaches to this problem. The method proposed by Maillet and collaborators \cite{Maillet:1985ek,Freidel:1991jx,Freidel:1991jv} seems, however, to be the simplest and most systematic in order to construct the action-angle variables, and understand the classical integrability. It involves a generalization of the concept of the $r$-matrix to a pair of $(r,s)$-matrices and the simultaneous regularization of the ill-defined Poisson brackets with the use of a symmetrization procedure to introduce the so-called Maillet brackets. The most fundamental difficulty that prevents the full implementation of the quantum inverse scattering method to non-ultralocal theories lies in finding the appropriate regularization/quantization of the corresponding Maillet brackets. In particular, one that recovers the symmetrization prescription pertaining the definition of the Maillet brackets in the classical limit. Thus, for the non-ultralocal systems there is no direct generalization of the Yang-Baxter type equation from which one can extract the quantum Hamiltonian and other quantum charges, and find the spectrum.

Although the lack of a general procedure to properly quantize the Maillet bracket has precluded the full implementation of the quantum inverse scattering method to many interesting models, one notable exception is the $SU(2)$ $PCM$ which has been quantized by Faddeev and Reshetikhin $(FR)$ in \cite{Faddeev:1985qu}. The $FR$ quantization method is based on the ultralocalization of the theory and its subsequent regularization in terms of a magnetic lattice algebra. The original non-ultralocality can be shown to be restored in  the continuous theory by taking the large spin limit. Thus, by replacing the original non-ultralocal Poisson algebra by a new ultralocal one, while preserving the equations of motion with respect to a new Hamiltonian, Faddeev and Reshetikhin avoided dealing directly with the problematic Maillet bracket. The recent identification of the algebraic mechanism underlying the above ultralocalization procedure enabled its application in more general contexts such as sigma models on symmetric and semi-symmetric spaces, including the $AdS_5\times S^5$ superstring \cite{Delduc:2012qb,Delduc:2012vq}. Notwithstanding all this effort the quantization of sigma models on symmetric and semi-symmetric spaces is still unknown, even though some candidate lattice Poisson algebras have been proposed \cite{Delduc:2012qb,Delduc:2012vq,Delduc:2012mk}.

From the few cases where the quantization of the Maillet bracket was successful \cite{Faddeev:1985qu,Alekseev:1992wn,Alekseev:1996jz,Freidel:1991jv} there emerges some general strategy to be followed. It reduces essentially to the following four steps: $(i)$ ultralocalize the Kac-Moody type algebra satisfied by the classical continuous theory; $(ii)$ regularize the ultralocalized current algebra to get rid of the singularities at coinciding points by invoking a lattice discretization; $(iii)$ quantize the lattice current algebra by means of the quantum inverse scattering method; $(iv)$ check that in the scaling limit the quantized discrete algebra reproduces the classical Kac-Moody algebra. However, this recipe breaks down for the $AAF$ model already in step $(i)$, as all the ultralocalization procedures so far developed work only for models plagued by non-ultralocalities up to the first derivative of the delta function, while the algebra of Lax operators for the $AAF$ model is even more non-ultralocal, including terms proportional to the second derivative of the delta function \cite{Melikyan:2012kj,Melikyan:2014yma}. Moreover, being a purely fermionic model, any naive lattice discretization necessarily incurs in fermion doubling.

Thus, despite the absence of appropriate methods to directly quantize the Maillet bracket for the $AAF$ model and the inherent difficulty in generalizing the available methods, one can still try to find the quantum Hamiltonian via coordinate Bethe Ansatz. For example, in the case of the $AAF$ model, it has been shown in \cite{Melikyan:2014mfa} that this is indeed possible, and the quantum Hamiltonian, which, after a field redefinition to make its Poisson structure canonical, acquires a very complex form, containing terms up to the eighth order in the fermion field and its derivatives, can be diagonalized. The key point here, as was shown in \cite{Melikyan:2014mfa}, is that the wave-functions (and their derivatives) in the quantum mechanical picture are not continuous functions, and to avoid meaningless expressions in the calculation one has to: (i)  treat the quantum fields as operator valued distributions, and (ii) employ the principal value prescription in the resulting integrations where the discontinuities arise. It was shown that this prescription indeed does the desired job, and the diagonalization process reproduces the correct $S$-matrix of the $AAF$ model, found earlier via perturbative calculations \cite{Klose:2006dd,Melikyan:2011uf}. 

In this paper we take another step towards the full implementation of the quantum inverse scattering method to the $AAF$ model. This program was initiated in \cite{Melikyan:2012kj} where we identified a surprisingly simple $2 \times 2$ representation for the Lax connection and showed that the resulting Poisson algebra was highly non-ultralocal. The second step, which entailed the development of an extension of the classical inverse scattering method and Maillet's $(r,s)$-formalism to accommodate the second derivative of the delta function in the algebra of Lax operators, was taken in \cite{Melikyan:2014yma}. Here we make the next step, and implement the principal value prescription, which had to be manually performed in the previous calculations, directly on the operator level. Namely, we show that by treating the quantum fields as operator valued distributions, and regularizing the product of operators by means of \emph{Sklyanin's product} \cite{Sklyanin:1988} in the quantum Hamiltonian, as well as any relevant operator quantities, such as the Lax operator, the principal value prescription follows naturally without any manual input. An immediate consequence of this implementation is the reproduction of Maillet's symmetrization prescription in the classical limit. We stress that differently from  \cite{Faddeev:1985qu,Freidel:1991jx,Freidel:1991jv,SemenovTianShansky:1995ha,Delduc:2012yz,Hollowood:2015dpa,Alekseev:1991wq,Alekseev:1992wn} we work directly in the continuous case, without appealing to any lattice regularization of the theory, as the latter is not always an easy task to formulate.

Sklyanin's product (or the $\circ$-product) is a type of split-point regularization, which was originally introduced in \cite{Sklyanin:1988} in order to regularize the product of two operators at the same point and therefore obtain the Yang-Baxter relation for the Landau-Lifshitz model. The latter is an ultralocal model, and so the difficulties of the quantization are associated only with the singularities appearing in the product of operators at the same point. Thus, the regularized quantum Hamiltonian can be naturally obtained from the fundamental regularized Yang-Baxter relations \cite{Melikyan:2008ab,Melikyan:2010fr}. In contrast, in the case of the $AAF$ model one does not have, as explained above, the Yang-Baxter type equations, and it is not clear from which fundamental relations such regularization with Sklyanin's product can appear. To address these points we consider the consistent reduction of the $AAF$ model to the free massive fermion  model. Such a procedure allows to avoid all the unnecessary technical complications of the $AAF$ model, and  automatically gives the Lax pair for the free fermion model which leads to an algebra with the same degree of non-ultralocality as the $AAF$, and thus it is still sufficiently non-trivial in order to test our approach. Then we show that if one regularizes the Lax operator via Sklyanin's product, the regularized quantum Hamiltonian can be obtained from the integral equations defining the quantum transition matrix. Moreover, for the classical theory obtained from this regularized quantum theory, one reproduces Maillet's symmetrized Poisson bracket prescription.

Our paper is organized as follows. In section \ref{sec:aaf} we present the most essential aspects of the $AAF$ model and briefly discuss how to generalize the classical inverse scattering method to accommodate its higher degree of non-ultralocality. Then, in section \ref{sec:sp}, we address the fundamental problem of ill-defined operator products when formulating a continuous quantum algebra and introduce Sklyanin's product as our regularizing prescription. Next, in section \ref{sec:fr_AAF}, we particularize the discussion of the previous section on the role of Sklyanin's product to the case of the $AAF$ model. In section \ref{sec:free}, we consider the reduction of the $AAF$ model to the free fermion model, which gives the explicit Lax operator and the associated non-ultralocal algebra. We explicitly work out the regularized quantum monodromy matrix, showing how to extract the quantum conserved charges. The relation between the normal product and Sklyanin's product is also explained. In section \ref{sec:qa}, building upon the results of the previous section, we conjecture the form of the quantum algebra of transition matrices for a non-ultralocal continuous theory and show that it consistently reduces to the Maillet algebra in the classical limit. In section \ref{sec:conclusion}, we summarize our results and point out some interesting directions and open problems. Finally, in appendices we collect various computational details used in the text.

\section{Overview of the Alday-Arutyunov-Frolov Model} \label{sec:aaf}

In this section we briefly overview the essential properties of the $AAF$ model, referring the reader to the papers \cite{Alday:2005jm,Arutyunov:2005hd,Melikyan:2011uf,Melikyan:2012kj,Melikyan:2014yma, Melikyan:2014mfa} for all the technical details. As we mentioned in the introduction the $AAF$ model arises from the reduction of the superstring on $AdS_{5} \times S^{5}$ to the $\mathfrak{su}(1|1)$ subsector, where in the process of constraint analysis all the bosonic degrees of freedom are eliminated in favor of fermionic ones. The resulting theory is a two-dimensional Lorentz-invariant fermionic model which is described by the following action (see appendix \ref{app:free_app} for notations):
\begin{align}
	 S &= \int dy^0 \: \int _0^J dy^1 \: \left[ i \bar{\psi} \delslash \psi \: - m \bar{\psi} \psi + \frac{g_2}{4m} \epsilon^{\alpha \beta} \left( \bar{\psi}
	\partial_{\alpha} \psi \; \bar{\psi}\: \gamma^3
	\partial_{\beta} \psi -
	\partial_{\alpha}\bar{\psi} \psi \;
	\partial_{\beta} \bar{\psi}\: \gamma^3 \psi \right) \right.- \nonumber \\
	&- \left. \frac{g_3}{16m} \epsilon^{\alpha \beta} \left(\bar{\psi}\psi\right)^2
	\partial_{\alpha}\bar{\psi}\:\gamma^3
	\partial_{\beta}\psi \right]. \label{aaf:action}
\end{align}
The two-particle scattering $S$-matrix has been first found from perturbative calculations and has the form \cite{Klose:2006dd}:
 \begin{equation}
S(\theta_{1},\theta_{2})=\frac{1-\frac{img_{2}}{4}\sinh(\theta_{1}-\theta_{2})}{1+\frac{img_{2}}{4}\sinh(\theta_{1}-\theta_{2})},\label{aaf:s_matrix}
 \end{equation}
where $\theta_{1}$ and $\theta_{2}$ are the rapidities of the scattered particles with momenta $p^{1}=m\sinh{\theta_{1}}$ and $p^{2}=m\sinh{\theta_{2}}$. The coupling constants $g_{2}$ and $g_{3}$ in \eqref{aaf:action} were introduced in \cite{Melikyan:2011uf}, where the $S$-matrix factorization property, underlying the quantum integrability of the model, was proved up to the first loop approximation, provided the relation $g_{2}^{2} = g_{3}$ between the coupling constants is satisfied.

The Lax pair for the $AAF$ model found in \cite{Alday:2005jm,Melikyan:2012kj} leads to a non-ultralocal algebra for the $L$-operators of the form \cite{Melikyan:2014yma}:\footnote{Here the symbol $\otimes$ stands for the \emph {supertensor product}, which extends the concept of the tensor product for bosonic fields to the fermionic case. For detailed mathematical definitions and the relevant constructions we refer the reader to the monograph \cite{Berezin:1987ue} and the original papers \cite{Kulish:1980ii,Kulish:1985bj,Gohmann:1998sm,Gohmann:1999av,Gohman:2002kp}. For a comprehensive review, see \cite{Essler:2005bk}.}

\begin{align}
	\{ {L}^{\scriptscriptstyle{(\sigma)}}(x;\lambda) \overset{\otimes}{,} {L}^{\scriptscriptstyle{(\sigma)}}(y;\mu) \} &= A(x,y;\lambda,\mu) \delta(x-y) +  B(x,y;\lambda,\mu) \partial_{x}\delta(x-y) \nonumber \\
	&+C(x,y;\lambda,\mu) \partial^{2}_{x}\delta(x-y) . \label{aaf:lax_general}
\end{align}
It has a more complicated form when compared to the standard case studied in \cite{Maillet:1985ek}, since it contains terms proportional not only to the first derivative of the delta-function, but also to its second derivative. To slightly simplify the discussion bellow, we shall classify the non-ultralocal algebras by the highest order of the derivative of the delta-function present. For instance, the algebra \eqref{aaf:lax_general} is a second order non-ultralocal algebra, while the standard case \cite{Maillet:1985ek}, a first order. Thus, to properly take into account the contribution of the last term in \eqref{aaf:lax_general} it is necessary to consider a generalization of Maillet's $(r,s)$-matrix formalism, which amounts to the introduction of a third matrix. In terms of the triple $(r,s_1,s_2)$, the algebra \eqref{aaf:lax_general} becomes:
\begin{align}\label{aaf:lax_algebra_r_s1_s2}
	\left\{ L^{\scriptscriptstyle{(\sigma)}}_1(z;\lambda)\right. &, \:\left. L^{\scriptscriptstyle{(\sigma)}}_2(z';\mu) \right\} = \delta (z-z')  \bigg( \partial_z r(z;\lambda,\mu) + \left[r(z;\lambda,\mu), L^{\scriptscriptstyle{(\sigma)}}_1(z;\lambda)  + L^{\scriptscriptstyle{(\sigma)}}_2(z;\mu) \right] \bigg. \\
	&+ \left[ s_1(z;\lambda,\mu), L^{\scriptscriptstyle{(\sigma)}}_2(z;\mu) - L^{\scriptscriptstyle{(\sigma)}}_1(z;\lambda) \right] + \left[ \partial_z s_2(z;\lambda,\mu), L^{\scriptscriptstyle{(\sigma)}}_1(z;\lambda) +  L^{\scriptscriptstyle{(\sigma)}}_2(z;\mu) \right] \notag \\
	&+ \left[ \left[ s_2(z;\lambda,\mu), L^{\scriptscriptstyle{(\sigma)}}_1(z;\lambda) \right], L^{\scriptscriptstyle{(\sigma)}}_2(z;\mu) \right] + \bigg. \left[ \left[ s_2(z;\lambda,\mu), L^{\scriptscriptstyle{(\sigma)}}_2(z;\mu) \right], L^{\scriptscriptstyle{(\sigma)}}_1(z;\lambda)  \right] \bigg) \: \notag \\
	&-  \partial_z \delta (z-z')\left[ s_1(z;\lambda,\mu) + s_1(z';\lambda,\mu)\right]  + \partial^2_z \delta (z-z') \left[ s_2(z;\lambda,\mu) + s_2(z';\lambda,\mu)\right]   \notag,
\end{align}
where we used the following standard notation for tensor products $L_{1}(z; \lambda) \equiv L(z; \lambda) \otimes \mathds{1}$ and $L_{2}(z;\lambda) \equiv \mathds{1} \otimes L(z;\lambda)$. For the $AAF$ model the exact form of the matrices $(r,s_{1}, s_{2})$ has a very complicated non-linear character \cite{Melikyan:2012kj}.

Nonetheless the resulting classical algebra of transition matrices corresponding to equal and adjacent intervals with $x>y>z$ has exactly the same structure as the originally proposed by Maillet \cite{Maillet:1985ek,Freidel:1991jx,Freidel:1991jv} for the simpler first order case,
\begin{align}
 \left\{ T_{1}(x,y;\lambda) , T_{2}(x,y;\mu) \right\}_{M} &=  r(x;\lambda,\mu) \: T_{1}(x,y;\lambda) T_{2}(x,y;\mu)- T_{1}(x,y;\lambda)  T_{2}(x,y;\mu) \: r(y;\lambda,\mu), \nonumber \\
	\left\{ T_{1}(x,y;\lambda) , T_{2}(y,z;\mu) \right\}_{M} &=  T_{1}(x,y;\lambda)  s(y;\lambda,\mu)  T_{2}(y,z;\mu). \label{aaf:T_algebra_symm}
\end{align}
The effect of the second derivative of the delta-function in \eqref{aaf:lax_algebra_r_s1_s2} amounts to the following shift of the pair of intertwining matrices $(r,s)$:
\begin{align}
	r(z;\lambda, \mu) \to u(z;\lambda, \mu) &= r(z;\lambda, \mu) + \partial_z s_2(z;\lambda,\mu) + \left[ s_2(z;\lambda,\mu), L_1(z;\lambda) + L_2(z;\mu)\right], \label{aaf:u-matrix}\\
	s(z;\lambda, \mu) \to v(z;\lambda,\mu) &= s_1(z;\lambda,\mu) + \left[ s_2(z;\lambda,\mu), L_1(z;\lambda) - L_2(z;\mu) \right]. \label{aaf:v-matrix}
\end{align}
In \eqref{aaf:T_algebra_symm} the subscript $M$ indicates that the Poisson bracket has to be symmetrized according to Maillet's prescription to avoid ambiguities arising from coinciding points. Starting from \eqref{aaf:T_algebra_symm} one can then construct the angle-action variables following the standard procedure \cite{Faddeev:1988qp,Novikov:1984id,Korepin:1997bk,Essler:2005bk}. This program has been realized for simpler models in \cite{Melikyan:2014yma}.

The fundamental construction underlying Maillet's method is the symmetrization prescription for Poisson brackets and the corresponding generalization for nested Poisson brackets. To introduce Maillet's symmetrization procedure one considers $n$-nested Poisson brackets for transition matrices $T(x_{i},y_{i};\lambda_{i})$:
\begin{align}
\Delta^{n}(x_{i},y_{i};\lambda_{i}) = \left\{ T(x_{1},y_{1};\lambda_{1}) \overset{\otimes}{,} \left\{ \ldots\overset{\otimes}{,}\left\{ T(x_{n},y_{n};\lambda_{n}) \overset{\otimes}{,} \;T(x_{n+1},y_{n+1};\lambda_{n}) \right\} \ldots \right\} \right\}, \label{aaf:nested_brackets}
\end{align}
and for any subset of $l=p+q$ coinciding points $x_{\alpha_{1}}= \ldots =x_{\alpha_p}=y_{\beta_{1}}= \ldots = y_{\beta_q}=z$, one defines the left-hand side of \eqref{aaf:nested_brackets} by:
\begin{align}
\Delta^{n}(z;\lambda_{i}) := \lim_{\epsilon \rightarrow 0} \frac{1}{l!} \sum_{\sigma  \,  {\scriptscriptstyle\in} \, \mathds{P}} \Delta^{n} \left(x_{\alpha_1} + \epsilon \sigma(1),\ldots,y_{\beta{q}} + \epsilon \sigma(l);\lambda_{i} \right), \label{aaf:symmetrization}
\end{align}
where for simplicity of notations we omitted in $\Delta^{n}(x_{i},y_{i};\lambda_{i})$ the dependence on the coordinates different from $z$, and the symbol $\mathds{P}$ indicates the sum over all possible permutations of $(1,\ldots,l)$. For example, this symmetrization procedure yields:
\begin{align}
	\{&T(x,y;\lambda) \overset{\otimes}{,} \;T(x,y';\mu) \}_{{M}} \nonumber\\
	& = \frac{1}{2} \lim_{\epsilon \rightarrow 0}  \left(\{T(x - \epsilon,y;\lambda) \overset{\otimes}{,} \;T(x + \epsilon,y';\mu) \} + \{T(x + \epsilon,y;\lambda) \overset{\otimes}{,} \;T(x - \epsilon,y';\mu) \} \right).\label{aaf:PB_sym}
\end{align}

The quantization of classical algebras for transition matrices of the form  \eqref{aaf:T_algebra_symm} has not been successful except in few very specific cases. One of the principal difficulties is that the commutators in the quantum theory cannot immediately reproduce the symmetrized Maillet brackets on the left hand side of \eqref{aaf:T_algebra_symm}. Another important difficulty is the quantization of integrable models directly in the continuous theory. This is especially relevant for the $AAF$ model as its lattice version is not known. Quantization of continuous models, as discussed in the introduction, presents a challenge due to the singularities arising from the product of operators. To correctly quantize the system, one should first remove such singularities by means of a proper regularization of the fields or products. Before addressing this problem for the $AAF$ model, we briefly explain in the next section how to quantize a simpler continuous integrable model - the Landau-Lifshitz ($LL$) model, which although ultralocal exhibits the same type of interaction terms in the Lagrangian as the $AAF$ model.

\section{Field regularization and operator product}
\label{sec:sp}

To formulate a well-defined algebra for quantum transition matrices for a continuous theory one must first deal with the singularities associated with operator products at the same point. The most natural way to solve this problem is to resort to the methods of quantum field theory where such singularities are dealt with by means of renormalization techniques. The latter can be strictly  formulated  in the framework of axiomatic quantum field theory (see, for example, the monograph \cite{streater2000pct}), where the quantum fields are treated as operator-valued distributions:
\begin{equation}
	\Phi_{\mathscr{F}}(x) = \int  dy\, \Phi(y)\mathscr{F}(x,y),\label{sp:field_distribution}
\end{equation}
where $\mathscr{F}(x,y) \equiv \mathscr{F}(x-y)$ is an element in the Schwartz space of test functions. 

The necessity to treat fields as distributions in the context of the integrable systems was first realized on the example of the $LL$ model in \cite{Melikyan:2008ab}, following an early attempt by Sklyanin \cite{Sklyanin:1988} to regularize the product of operators in order to satisfy the Yang-Baxter relation.\footnote{In the original Sklyanin's approach \cite{Sklyanin:1988}, instead of treating fields as distributions,  a product between two operators was introduced in order to to regularize arising singularities. The original definition of \cite{Sklyanin:1988} is as follows:
\begin{equation}
	\label{sp:sklyanin_product_original} A(x) \circ B(x) \equiv \lim_{\Delta \to 0} \frac{1}{\Delta} \int _x^{x+\Delta} d\xi_1 \: \int _x^{x+\Delta} d\xi_2 \: A(\xi_1) B(\xi_2).
\end{equation}
Although this was enough to obtain the Yang-Baxter equation \eqref{sp:Yang_Baxter}, Sklyanin's product, as  discussed in \cite{Melikyan:2010fr}, was not enough to diagonalize the quantum Hamiltonian, or to obtain the higher order local conserved charges. It also led to other problematic singular expressions.} More recently this approach was also applied to the $AAF$ model \cite{Melikyan:2014mfa}. For both models it was shown that in order to achieve exact diagonalization of the quantum Hamiltonian and to construct the $n$-particle sector wave-functions,  one has to regularize the fields as in  \eqref{sp:field_distribution}. This allows to avoid meaningless singularities of the type $~\partial_{x}^{2}\delta(0)$, and permits  the construction of the correct spectrum and $S$-matrix. Furthermore, it was also shown that the corresponding quantum-mechanical Hamiltonian is a self-adjoint operator. These results would have been impossible to obtain without regularizing the fields as in \eqref{sp:field_distribution}, i.e., treating the fields as distributions. Before turning to the more complex, non-ultralocal integrable $AAF$ model, we first give a brief review of this construction for the simpler, ultralocal $LL$ model using the methods previously elaborated in \cite{Sklyanin:1988,Melikyan:2008ab,Melikyan:2010fr}. Then we present a new, more convenient formulation, which is more appropriate when dealing with non-ultralocal models. 

We recall, that the Hamiltonian for the isotropic $LL$ model for the $\mathfrak{su}(1,1)$ case has the form \cite{Faddeev:1987ph}:
\begin{equation}
	H = \frac{1}{2} \int dx \: \left( \partial_x {\vec{S}}	\cdot \partial_x {\vec{S}} \right) ,\label{sp:ll_hamiltonian}
\end{equation}
where  the fields $S^{i}$ $(i =1,2,3)$ satisfy the following Poisson structure:\footnote{Here $S^{\pm }=S^{1}\pm iS^{2}$.}
\begin{align}
	\left\{ S^{3}(x),S^{\pm }(y)\right\} & =\pm iS^{\pm }(x)\delta (x-y),  \label{sp:poissonstructure} \\
	\left\{ S^{-}(x),S^{+}(y)\right\} & =2i S^{3}(x)\delta (x-y). \notag
\end{align}
It is now possible to show (for full details see \cite{Melikyan:2010fr}) that by passing to regularized fields as in \eqref{sp:field_distribution}:
\begin{equation}
S^{i}_{\mathscr{F}}(x) = \int dy \, S^{i}(y) \mathscr{F}(x,y),\label{sp:S_field_distribution}
\end{equation}
the Lax operator:
\begin{equation}
	\label{sp:LF_operator} \mathcal{L}^{\mathscr{F}}(\lambda,x) = \frac{i}{\lambda}\left(
	\begin{array}{cc}
		S^3_{\mathscr{F}}(x) & -S^+_{\mathscr{F}}(x) \\
		S^-_{\mathscr{F}}(x) & -S^3_{\mathscr{F}}(x)
	\end{array}
	\right)
\end{equation}
satisfies the fundamental intertwining relation:
\begin{align}
	\lim \left\{ R(\lambda_1-\lambda_2) \left[ { \mathcal{L}}_1^{\mathscr{F}}(\lambda_1,x) + {\mathcal{L}}_2^{ \mathscr{F}}(\lambda_2,x) + {\mathcal{L}}_1^{\mathscr{F} }(\lambda_1,x) \cdot {\mathcal{L}}_2^{\mathscr{F}}(\lambda_2,x) \right] \right\} = \notag \\
	= \lim \left\{ \left[ {\mathcal{L}}_1^{\mathscr{ F }}(\lambda_1,x) + {\mathcal{L}}_2^{\mathscr{F}}(\lambda_2,x) + {\mathcal{L}}_2^{\mathscr{F}}(\lambda_2,x) \cdot { \mathcal{L}}_1^{\mathscr{F}}(\lambda_1,x) \right] R(\lambda_1-\lambda_2) \right\}. \label{sp:bilinearrelationforLF}
\end{align}
The limit on both sides of the equation \eqref{sp:bilinearrelationforLF} corresponds to removing the regularization, i.e., when ${\mathscr{F}}(x) \sim \delta(x)$. The quantum $R(\lambda)$-matrix in the above expression is given by the following formula:
\begin{equation}
	\label{sp:R_matrix} R(\lambda) = \sum^{3}_{a=0} w_a(\lambda) \sigma_a \otimes \sigma_a,
\end{equation}
where $w_0(\lambda) = \lambda - \nicefrac{i}{2}$, $w_{1,2,3} = - \nicefrac{i}{2}$, and $\sigma_{a}=({\mathds{1},\sigma_{i}})$.

Furthermore, it can be shown \cite{Sklyanin:1988} that the intertwining relation  \eqref{sp:bilinearrelationforLF} leads to the following Yang-Baxter relation:
\begin{equation}
	 R(\lambda -\mu) {T}_1(\lambda) {T}_2(\mu) ={T}_1(\mu) {T}_2(\lambda) R(\lambda-\mu), \label{sp:Yang_Baxter}
\end{equation}
where  the monodromy matrix $T(\lambda)$ is obtained from the corresponding quantum $ \mathcal{L}^{\mathscr{F}}(\lambda,x)$. The Yang-Baxter relation  \eqref{sp:Yang_Baxter}  allows the quantization of the $LL$ model using the standard methods. In particular, using the regularized fields as discussed above, one can diagonalize the quantum Hamiltonian for any $n$-particle sector, as well as construct the higher order conserved charges. 

%In what follows we give a generalisation of the Sklaynin product %\eqref{sp:sklyanin_product_original}, and show that it can be obtained from the  %fields treated as distributions.

In order to present our results, it is necessary to first explain how the diagonalization procedure for the LL model should be carried out when the fields are regularized according to \eqref{sp:S_field_distribution}. Afterwards, we introduce an alternative formulation which is more suitable for non-ultralocal models such as the $AAF$ model. The main result of  \cite{Melikyan:2010fr} is that  the quantum Hamiltonian of the $LL$ model written in terms of the $\mathscr{F}$-regularized fields has the form:
\begin{equation}
	\label{sp:fregularizedhamiltonian} \mathcal{H}_{\mathscr{F}} = \frac{1}{4} \int dx \: \left[ -2
	\partial_x S^3_{\mathscr{F} }(x)
	\partial_x S^3_{\mathscr{F}}(x) +
	\partial_x S^+_{\mathscr{F}}(x)
	\partial_x S^-_{\mathscr{F}}(x) +
	\partial_x S^-_{\mathscr{F}}(x)
	\partial_x S^+_{\mathscr{F}}(x) \right],
\end{equation}
while the $n$-particle states are:
\begin{equation}
	\label{sp:fntilde} |f_n\rangle = \int \prod_{i=1}^n dx_i \: f _n(x_1,\ldots, x_n ) \prod_{j=1}^n S^+(x_j) |0\rangle,
\end{equation}
and provide a representation space for the $\mathfrak{su}(1,1)$ algebra for the operators in terms of $S^{i}_{\mathscr{F}}$ fields. Here, the wave functions $f_n(x_1,\ldots,x_n)$ can be shown to be continuous and sufficiently fast decreasing, symmetric functions of $x_{1} ,\ldots, x_{n}$, which, however, have discontinuous first derivatives. 

The crucial comment is that due to the presence of the derivatives in the quantum Hamiltonian \eqref{sp:fregularizedhamiltonian}, and the discontinuity of the first derivatives of the wave functions, during the process of the diagonalization, the resulting integrations should be understood in the \emph{principal value} sense. In other words, even though the field regularization \eqref{sp:S_field_distribution} is enough to obtain the Yang-Baxter relation \eqref{sp:Yang_Baxter}, one still has to treat the integrals arising in the diagonalization process in the principal value sense. As was shown in  \cite{Melikyan:2010fr} this leads to the boundary conditions on the first derivatives of $f_n(x_1,\ldots,x_n)$, and indeed reproduces the correct $S$-matrix. Thus, whenever integrals containing the derivatives $\partial_{x_{i}}f_n(x_1,\ldots,x_n)$ occur, one has to understand such integrals in the principal value sense. For example, for the case  $n=2$,  the arising  integrals are of  the type:
\begin{equation}
\iint dx\, dy \,\, \partial_{x}^{l}\partial_{y}^{k}\left[\partial_{x}f_{2}(x,y)\ldots \right];\,\,\,\,\qquad l,k=0,1,\ldots, \notag
\end{equation}
and should, in general, be understood as follows:
\begin{equation}
\int dx\, \left[ \int_{-\infty}^{x-\varepsilon} dy + \int_{x+\varepsilon}^{\infty} dy \;\right]  \partial_{x}^{l}\partial_{y}^{k}\left[\partial_{x}f_{2}(x,y)\ldots \right]. \notag
\end{equation}
For the  higher $n$-particle sectors,  the resulting integrations are over an $n$-dimensional space, and one has to divide the integration region into $n!$ subspaces, similar to the example above, in such a way as to remove a small strip around the singularity region.

Thus, for a general model, when performing the direct diagonalization one has to manually take care of ill-defined integrals due to discontinuities of the wave functions or their derivatives. The natural question is whether it is possible to have a more fundamental formulation which does not require any such manual input, and the principal value prescription or its $n$-dimensional generalization is taken into account from the beginning. We show next that this is indeed possible, and can be formulated in terms of an operator product. For two operators $A(x)$ and $B(x)$, the $\circ$-product is defined  as follows:\footnote{We also refer to it as \emph{Sklyanin's product}, since it can be seen as a modification of Sklyanin's original definition \eqref{sp:sklyanin_product_original}.}
\begin{equation}
A(x) \circ B(x)
\equiv  \lim_{\nicefrac{\Delta}{2} \to \epsilon} \frac{1}{(\Delta - \epsilon)^{2}} \fint\limits_{\Delta S_{1} \cup \Delta S_{2}}d\zeta d\xi \;A(\zeta)B(\xi). \label{sp:sklyanin_prod}
\end{equation}
The notation $\fint\limits_{\Delta S_{1} \cup \Delta S_{2}}$ means that the integration is taken over a square of side $\Delta$, minus a strip of width $\epsilon$ around the diagonal $\zeta=\xi$, and the areas $\Delta S_{1}$ and $\Delta S_{2}$ correspond to the regions above the line $\zeta=\xi + \epsilon$ and below the line $\zeta = \xi - \epsilon$. This essentially means that we ``smear'' the product of two operators around an arbitrary small area of size $\Delta$, avoiding the singularity at $\zeta=\xi$. The parameter $\epsilon$ is the regularization parameter of the theory, and should be taken to zero only at the end of all computations. It is clear from \eqref{sp:sklyanin_prod} that if the product of two operators is not singular, then, in the limit $\epsilon \to 0$, it reduces to the usual product. In other words, we explicitly exclude the entire singular region, parametrized by the length scale $\epsilon$.

In general, for a product of $k$ operators $A_{1}(x), \ldots, A_{k}(x)$, the $\circ$-product is defined similarly to \eqref{sp:sklyanin_prod}. The integration should now be performed over the $k$-dimensional cube of side $\Delta$, where all possible singular regions are taken out of the integration domain. Thus, there are $k!$ integrations over disconnected volume elements of size $\Delta V_{i}$ corresponding to all possible orderings of the variables $\zeta_{1}, \ldots, \zeta_{k}$ separated by the length of the regularization parameter $\epsilon$. More precisely, we have:
\begin{equation}
A_{1}(x) \circ \ldots  \circ A_{k}(x)
\equiv  \lim_{\nicefrac{\Delta}{2} \to \epsilon} \frac{1}{\Delta V} \sum\limits_{i=1}^{k!}\int\limits_{\Delta V_{i}} d\zeta_{1} \ldots d\zeta_{k} \;  A_{1}(\zeta_{1}) \cdot \cdots \cdot A_{k}(\zeta_{k}), \label{sp:sklyanin_prod_general}
\end{equation}
where $\Delta V$ is the sum of all disconnected volume elements $\Delta V_{i}$.

\subsection*{Principal value prescription from operator product}

We now state one of our central results. In order to account for the principal value prescription from the beginning, which is needed, as discussed above, to treat ill-defined integrals due to the discontinuities of the wave functions $f_n(x_1,\ldots,x_n)$ or their derivatives, one has to replace the usual operator product in the quantum Hamiltonian (or in other relevant operators, e.g., higher-dimensional charges) by the $\circ$-product defined in \eqref{sp:sklyanin_prod_general}. We stress that this is in addition to regularizing the fields as in \eqref{sp:field_distribution}. To prove this statement we explicitly  consider, for the the sake of concreteness,\footnote{We note, nonetheless, that a similar situation takes place in other integrable models, including the $AAF$ model, and, therefore, the proof that follows can be easily generalized for other types of interactions, with suitable modifications.} the action of the following term from the Hamiltonian of the $LL$ model \eqref{sp:fregularizedhamiltonian}:
\begin{align}\label{sp:ll_hamiltonian_term}
\mathcal{H}_{\mathscr{F}} = \int dx \: \partial_x S^3_{\mathscr{F}}(x) \partial_x S^3_{\mathscr{F}}(x)
\end{align}
on the two-particle state \eqref{sp:fntilde}:
\begin{align}\label{sp:2p_state}
	|f_2\rangle = \int d\xi \: d\zeta \: f_2(\xi,\zeta) S^+(\xi) S^+(\zeta) |0\rangle.
\end{align}
The result reads:
\begin{align}
	\mathcal{H}_{\mathscr{F}}|f_2\rangle= \int dx \; du \; dv \; f_{2}(u,v)\partial_{u} \mathscr{F}(x,u)\partial_{v} \mathscr{F}(x,v) S^{+}(u) S^{+}(v)|0\rangle + \ldots, \label{sp:I_tilde_original}
\end{align}
where the terms in ellipses are suppressed in order to avoid cluttering, and have a similar to the first term structure. To evaluate such terms and to obtain meaningful expressions, one must employ the principal value prescription, as discussed earlier. After some transformations one arrives at the following result \cite{Melikyan:2010fr}:\footnote{Here $\delta_{\alpha}(x)$ denotes a regularization of the $\delta$-function with respect to the regularizing length parameter $L=\nicefrac{1}{\alpha}$ \cite{Weinberg1995:qf1}:
\begin{align}
	\delta_L(x) = \frac{1}{2 \pi}\int_{-L/2}^{L/2}  e^{ix u} \; du.
\end{align}
The length parameter $L$  is very large and is associated with the size of the box in which we consider our system, resulting, as a consequence, in asymptotic Bethe ansatz equations.}
\begin{align}
	\mathcal{H}_{\mathscr{F}}|f_2\rangle &= \int dx \; du \; \left[\int_{-\infty}^{u-\epsilon} dv + \int_{u+\epsilon}^{\infty} dv \right]
	\left\{ 
	 \partial_{v}\partial_{u}f_{2}(u,v) \mathscr{F}(x,u)\mathscr{F}(x,v)S^{+}(u)S^{+}(v) \right. \notag \\
	& \left. + \; \partial_{u}f_{2}(u,v)\mathscr{F}(x,u)\mathscr{F}(x,v)S^{+}(u)\partial_{v}S^{+}(v)   \right\} \vert 0 \rangle \notag \\
	&- \frac{1}{\delta_{\alpha}(0)} \int du \; \partial_{u}f_{2}(u,v)\vert^{v=u-\epsilon}_{v=u+\epsilon} \; S^{+}(u)S^{+}(v)\vert 0 \rangle +\ldots \;. \label{sp:I_tilde}
\end{align}
Next, we show that by employing the operator product \eqref{sp:sklyanin_prod}
one arrives at the same result \eqref{sp:I_tilde}, but without the necessity to use any principal value prescription. In other words, we show that the principal value prescription can be implemented on the operator level. The advantage of this will be discussed further in the text.

We start by writing the Hamiltonian \eqref{sp:ll_hamiltonian_term} in terms of the $\mathscr{F}$-regularized fields and the operator product \eqref{sp:sklyanin_prod}:
\begin{align}\label{sp:ll_hamiltonian_term_f_reg_sp}
\mathcal{H}_{\mathscr{F}}^{SP} &= \int dx \; \partial_x S^3_{\mathscr{F}}(x) \circ \partial_x S^3_{\mathscr{F}}(x)\\
&= \lim_{\Delta \to 2\epsilon} \frac{1}{(\Delta - \epsilon)^2} \int dx \: dy \: dz \: \left[ \int\limits_{x-\frac{\Delta}{2}+\epsilon}^{x+\frac{\Delta}{2}}du \int\limits_{x-\frac{\Delta}{2}}^{u-\epsilon}dv\;  + \int\limits_{x-\frac{\Delta}{2}}^{x+\frac{\Delta}{2} - \epsilon} du \int\limits_{u+\epsilon}^{x+\frac{\Delta}{2}} dv \right] \nonumber \\
	&\cdot \partial_u \mathscr{F}(u,y) \partial_v \mathscr{F}(v,y) S^3(y) S^3(z). \nonumber
\end{align}
Using the commutation relations following from the Poisson algebra \eqref{sp:poissonstructure} and that $\mathscr{F}(x,y)$ depends only on the difference of its arguments \cite{Melikyan:2010fr}, it is a straightforward computation to show that the action of \eqref{sp:ll_hamiltonian_term_f_reg_sp} on \eqref{sp:2p_state} involves the evaluation of integrals of the type bellow:
\begin{align}
 \mathcal{H}_{\mathscr{F}}^{SP}|f_2\rangle &= \lim_{\Delta \to 2\epsilon} \frac{1}{(\Delta - \epsilon)^2}\int dx \: dy \: dz \: \Biggl\{ \left[ \int\limits_{x-\frac{\Delta}{2}+\epsilon}^{x+\frac{\Delta}{2}}du \int\limits_{x-\frac{\Delta}{2}}^{u-\epsilon}dv\;  + \int\limits_{x-\frac{\Delta}{2}}^{x+\frac{\Delta}{2} - \epsilon} du \int\limits_{u+\epsilon}^{x+\frac{\Delta}{2}} dv \right] \nonumber \\ 
& \cdot f_2(y,z) \partial_y \mathscr{F}(u,y) \partial_z \mathscr{F}(v,z) S^+(y) S^+(z) \Biggl\} |0\rangle+ \ldots \;. \label{sp:sample_integral}
\end{align}
Here we have explicitly written down the term corresponding to the first term in \eqref{sp:I_tilde_original}, while the other terms represented by ellipses are again suppressed. The $\circ$-product splits the integration into two disconnected regions which differ only on the domain of integration for the variables $u$ and $v$, as one can clearly see in \eqref{sp:sample_integral}. Thus, in the following we concentrate on the evaluation of 
\begin{align}
	I_a &= \int dx \: dy \: dz \int\limits_{x-\frac{\Delta}{2}+\epsilon}^{x+\frac{\Delta}{2}}du \int\limits_{x-\frac{\Delta}{2}}^{u-\epsilon}dv \: \partial_y\mathscr{F}(u,y) \partial_z \mathscr{F}(v,z)  f_2(y,z) S^+(y) S^+(z),\label{sp:Ia}
\end{align}
noting only that the computation of 
\begin{align}
	I_b &= \int dx \: dy \: dz \int\limits_{x-\frac{\Delta}{2}}^{x+\frac{\Delta}{2} - \epsilon} du \int\limits_{u+\epsilon}^{x+\frac{\Delta}{2}} dv \: \partial_y\mathscr{F}(u,y) \partial_z \mathscr{F}(v,z) f_2(y,z) S^+(y) S^+(z) \label{sp:Ib}
\end{align}
is completely analogous.

Since for the $LL$ model the wave function $f_2(y,z)$ is itself continuous, all the functions being integrated in $I_a$ \eqref{sp:Ia} are continuous in the region above the line $u=v$, so that we can invoke the mean value theorem to compute the integrals over $u$ and $v$. This makes $u=x+c$ and $v=x-k$, where $c,k \in \left( \epsilon, \frac{\Delta}{2} \right)$, and \eqref{sp:Ia} reduces to:
\begin{align}\label{sp:Ia_after_MVT}
	I_a = \frac{(\Delta - \epsilon)^2}{2} \int dx \: dy \: dz \: f_2(y,z) \partial_y \mathscr{F}(x+c,y) \partial_z \mathscr{F}(x-k,z) S^+(y) S^+(z).
\end{align}
Next, we integrate by parts with respect to $y$ and use the fact that the fields $S^+(x)$ vanish as $|x|\to \infty$ to obtain:
\begin{align}\label{sp:Ia_parts_y}
	I_a = -  \frac{(\Delta - \epsilon)^2}{2} \int dx \: dy \: dz &\Big\{ \partial_y f_2(y,z) \mathscr{F}(x+c,y) \partial_z \mathscr{F}(x-k,z) S^+(y) S^+(z) \Big.  \\
	&+ \Big. f_2(y,z) \mathscr{F}(x+c,y) \partial_z \mathscr{F}(x-k,z) \partial_y S^+(y) S^+(z)\Big\}. \nonumber
\end{align}
The first term in \eqref{sp:Ia_parts_y} is proportional to the derivative of the wave function, $\partial_y f_2(y,z)$, which is no longer continuous on the line $y=z$ for the $LL$ model. Thus, when evaluating \eqref{sp:Ia_parts_y} one needs to carefully consider the first term, while the second term can be trivially integrated by parts with respect to $z$. Hereafter, we will drop the contribution from the second term in \eqref{sp:Ia_parts_y}. We note, nonetheless, that the contribution from the corresponding term has also been neglected in the computation that led to \eqref{sp:I_tilde}.

The principal value prescription was introduced in \eqref{sp:I_tilde} with the sole purpose of avoiding the discontinuity of the derivative of the wave function on the line $y=z$. Here, we will show that the $\circ$-product, by dividing the integration domain over $u$ and $v$ into two disjoint regions, naturally avoids such discontinuity. In order to do that we will explicitly use the fact that the fields $S_{\mathscr{F}}^{+}(x)$  are smeared around the point $x$, i.e., they are localized in a small, but finite, neighborhood of the point $x$. The key point being that such neighborhoods can be taken sufficiently small to be completely separated by the width $\epsilon$ introduced by Sklyanin's product. Since, in the end of the calculation, we are going to remove the $\mathscr{F}$-regularization, it is sufficient to show that there is a representation of the $\mathscr{F}$-functions in terms of delta sequences that satisfies such separation property. For the sake of concreteness, we consider  the following explicit representation \cite{Vladimirov:2002mt}: 
\begin{align}\label{sp:F-function_rep}
	\mathscr{F}_{\alpha}(x,y) = 
  \begin{cases} 
   C_{\alpha} e^{-\frac{\alpha^2}{\alpha^2-|x-y|^2}} & \text{if } |x-y| \leq \alpha \\
   0       & \text{if } |x-y| > \alpha,
  \end{cases}
\end{align}
where $C_{\alpha}$ are constants satisfying:
\begin{equation}
	\int dz\: \mathscr{F}_{\alpha}(z) =1, \quad \Longleftrightarrow \quad C_{\alpha}\alpha^{n} \int_{|z|<1}dz \; e^{-\frac{1}{1-|z|^2}}=1.
\end{equation}
Here we denoted $\mathscr{F}_{\alpha}(x-y) \equiv \mathscr{F}_{\alpha}(x,y)$.

It can be proved that provided the regulator $\alpha$ is kept finite the representation \eqref{sp:F-function_rep} of $\mathscr{F}_{\alpha}(x,y)$ and its derivatives are smooth functions with support in $|x-y|\leq \alpha$. Thus, the integrations over $y$ and $z$ in \eqref{sp:Ia_parts_y} have only non-zero contribution from the intervals:
\begin{align}
	x + c - \alpha_1 \leq y \leq x + c + \alpha_1, \label{sp:ie_y}\\
	x - k - \alpha_2 \leq z \leq x - k + \alpha_2. \label{sp:ie_z}
\end{align}
Noting that $c,k \in  \left( \epsilon, \frac{\Delta}{2} \right)$, it is easy to see that whenever $\alpha_1, \alpha_2 <\epsilon$ the following inequalities hold
\begin{align}\label{sp:ie_yz}
	\alpha_1 + \alpha_2 < c + k \: \Leftrightarrow \:z \leq x - k + \alpha_2 < x + c - \alpha_1 \leq y
\end{align}
and the line $y=z$ is never reached in \eqref{sp:Ia_parts_y}. In particular, if we choose $\alpha_1,\alpha_2 < \frac{\epsilon}{2}$, the inequalities \eqref{sp:ie_y} and \eqref{sp:ie_z} together with the fact that $c,k > \epsilon$ restrict the domain of integration over $z$ to $(-\infty, y -\epsilon)$. The choice of $\alpha_1,\alpha_2$ to be smaller than $\epsilon$ can be justified as follows. Recall, that $\epsilon$ is the characteristic length around the singularity region, which is removed in the definition of the $\circ$-product \eqref{sp:sklyanin_prod}. In other words, the inverse $\Lambda=\nicefrac{1}{\epsilon}$ can be interpreted as the cut-off length, which is a fixed momentum and remains constant throughout the calculation. Any other parameters should be taken to zero before considering (if necessary) the limit $\epsilon \to 0$.

Therefore, after integrating by parts with respect to $z$, \eqref{sp:Ia_parts_y} becomes:\footnote{Here, $\partial_j f$ denotes the derivative with respect to the j-th argument of $f$.}
\begin{align}\label{sp:Ia_parts_z}
	I_a =  -  \frac{(\Delta - \epsilon)^2}{2} \int &dx \: dy \: dz \Big\{ \partial_1 f_2(y,y-\epsilon) \mathscr{F}_{\alpha_1}(x+c,y) \mathscr{F}_{\alpha_2}(x-k,y-\epsilon) S^+(y) S^+(y-\epsilon) \Big. \nonumber \\ \nonumber
	&- \partial_z \partial_y f_2(y,z) \mathscr{F}_{\alpha_1}(x+c,y) \mathscr{F}_{\alpha_2}(x-k,z) S^+(z) S^+(y) \\
	&- \Big.\partial_y f_2(y,z) \mathscr{F}_{\alpha_1}(x+c,y) \mathscr{F}_{\alpha_2}(x-k,z) S^+(y) \partial_zS^+(z) \Big\} + \cdots. 
\end{align}

Dividing \eqref{sp:Ia_parts_z} by $\left( \Delta - \epsilon\right)^2$ and taking the limit $\Delta \to 2\epsilon$ forces the open interval $\left( \epsilon, \frac{\Delta}{2} \right)$ to shrink to $\epsilon$. Finally, in order to remove the $\mathscr{F}$-regularization, we can use the relation between $\mathscr{F}$-functions and delta sequences \cite{Melikyan:2010fr}. Hence, after considering the appropriate limit, we can perform the integration over $x$ to obtain:
\begin{align} \label{sp:Ia_final}
	\lim_{\Delta \to 2 \epsilon} \frac{I_a}{(\Delta - \epsilon)^2}& = -\frac{1}{2\delta_{\alpha}(0)} \int dy\:  \partial_1 f_2(y,y-\epsilon) S^+(y) S^+(y-\epsilon) \\
	&+\frac{1}{2(\delta_{\alpha}(0))^2} \int dy\: \left[ \partial_2 \partial_1 f_2(y+\epsilon, y-\epsilon) S^+(y+\epsilon) S^+(y-\epsilon) \right. \nonumber \\
	&+ \left. \partial_1f_2(y+\epsilon,y-\epsilon) S^+(y+\epsilon) \partial_1S^+(y-\epsilon) \right] + \cdots . \nonumber
\end{align}
Proceeding analogously with \eqref{sp:Ib}, we derive a very similar expression for the contribution from bellow the line $u=v$. Summing these two results, we finally obtain:
\begin{align}\label{sp:I_final}
	&\mathcal{H}_{\mathscr{F}}^{SP}|f_2\rangle = \lim_{\Delta \to 2 \epsilon} \frac{I_a + I_b}{(\Delta - \epsilon)^2} |0\rangle + \cdots \\
	 &= \frac{1}{2\delta_{\alpha}(0)} \int dy\left[  \partial_1 f_2(y,y+\epsilon) S^+(y+\epsilon) - \partial_1f_2(y,y-\epsilon)  S^+(y-\epsilon) \right] S^+(y) |0\rangle \nonumber \\
	&+ \frac{1}{2(\delta_{\alpha}(0))^2} \int dy\: \Big\{ \left[ \partial_2 \partial_1 f_2(y+\epsilon, y-\epsilon) + \partial_2 \partial_1 f_2(y-\epsilon, y+\epsilon) \right] S^+(y+\epsilon) S^+(y-\epsilon) \Big. \nonumber \\
	& + \Big. \partial_1f_2(y+\epsilon,y-\epsilon) S^+(y+\epsilon) \partial_1S^+(y-\epsilon) + \partial_1f_2(y-\epsilon,y+\epsilon) S^+(y-\epsilon) \partial_1S^+(y+\epsilon) \Big\} |0\rangle + \cdots, \nonumber
\end{align}
which reproduces \eqref{sp:I_tilde} without the need to invoke any principal value prescription in the middle of the calculation. 

Hence, we see that Sklyanin's product \eqref{sp:sklyanin_prod_general} naturally avoids the discontinuities in the derivatives of the wave functions both in the boundary and bulk terms. Moreover, all the operator products in \eqref{sp:I_final} are explicitly symmetrized, indicating that the $\circ$-product \eqref{sp:sklyanin_prod} provides a quantization which is compatible with Maillet's symmetrization prescription \eqref{aaf:PB_sym} in the classical case. Later, in section \ref{sec:qa}, we will work out in details the relation between the quantum regularization provided by the $\circ$-product and the classical regularization given by Maillet's brackets. 

We conclude this section with remarks on the physical meaning of the $\circ$-product \eqref{sp:sklyanin_prod}. It follows directly from the definition that this  product  is a point splitting type of regularization, which is frequently used in quantum field theory in order to regularize singular operator products, e.g., in relation to anomaly computations (for an overview and the relation of point-splitting regularization to other methods, see \cite{Novotny:1998bw,Moretti:1998rf}). We recall the standard example of the axial fermionic electromagnetic current, regularized via point splitting method \cite{Peskin:1995qft}:
\begin{align}
	j_{5}^{reg\,\, \mu}(x) = \bar{\psi}(x+\nicefrac{\eta}{2}) \gamma^{\mu}\gamma_{5}e^{-ie\int_{x-\eta/2}^{x+\eta/2}dy^{\nu}A_{\nu}}\psi(x-\nicefrac{\eta}{2}).\label{sp:anomaly}
\end{align}
To avoid the dependence on the fictitious point $\eta^{\mu}$, one has to take the limit $\eta^{\mu} \to 0$ \emph{symmetrically}, which means that it should be taken in such a way as to ensure that the final formulas do not depend on the specifically chosen vector $\eta^{\mu}$. More explicitly, the symmetric limit is defined as follows: (for details see \cite{Peskin:1995qft}):
\begin{align}
	\text{symm} \lim_{\eta \to 0} \left[ \frac{\eta^{\mu}}{\eta^{2}} \right] &=0 \label{sp:sym1}\\
	\text{symm} \lim_{\eta \to 0} \left[ \frac{\eta^{\mu}\eta^{\nu}}{\eta^{2}} \right] &= \frac{1}{2} g^{\mu \nu} \label{sp:sym2}.
\end{align}
In contrast, the product in \eqref{sp:sklyanin_prod} is automatically smeared homogeneously over the entire region around the singularity strip, and does not depend on any chosen point $\eta^{\mu}$.

To summarize, in order to derive the standard intertwining relation \eqref{sp:bilinearrelationforLF} for the $LL$ model and perform diagonalization of the quantum charges one has to treat quantum fields as operator valued distributions \eqref{sp:field_distribution}, and, in addition, employ Sklyanin's product \eqref{sp:sklyanin_prod}  to account in a natural way for principal value prescription when dealing with discontinuous wave-functions or their derivatives. One of the consequences of our regularization is that it reproduces Maillet's symmetrization prescription for non-ultralocal systems in the classical limit. Further implications of this new formulation will be considered in the next section for the $AAF$ model.

\section{Field regularization for the $AAF$ model}
\label{sec:fr_AAF}
We now turn our attention to the $AAF$ model, which unlike the $LL$ model, is not an ultralocal integrable model. The immediate difficulty in this case is the following. For the ultralocal models, such as the $LL$ model, the quantum Hamiltonian as well as the other conserved charges can be found from the fundamental Yang-Baxter relation \eqref{sp:Yang_Baxter}. In contrast, for non-ultralocal models, such as the $AAF$ model, there does not exist such quantum relation, from which one can, for example, extract the quantum Hamiltonian. We emphasize, that such quantum Hamiltonian should be written, according to our main result in the previous section, in terms of the regularized fields \eqref{sp:field_distribution}, where the operator products are written in terms of Sklyanin's product \eqref{sp:sklyanin_prod}, in order to avoid the problems associated with singularities and to implement the principal value prescription from the beginning.

The $AAF$ model \eqref{aaf:action} presents further complications in comparison to simpler models. Namely, the Dirac brackets for the fermionic fields have a very complex structure, extending up to the eighth order in the fields and their derivatives. This immediately creates a computational difficulty when dealing with regularized fields \eqref{sp:field_distribution}. An alternative approach was taken in \cite{Melikyan:2014mfa} where the so-called equivalence theorem  for field theories \cite{Bergere:1975tr} was proven for the $AAF$ model. The equivalence theorem states (see \cite{Bergere:1975tr} and the references therein) that the $n$-particle $S$-matrix of the theory, which for integrable models plays a central role and can be used to reconstruct the spectrum of the model, does not change under an appropriate transformation of the quantum fields in the action \eqref{aaf:action}. 

One such change of fields, which was studied in details in \cite{Melikyan:2014mfa}, results in the theory for which the complicated non-linear Dirac brackets of the original theory \eqref{aaf:action} are reduced to the canonical relations. This in turn allows the possibility of directly diagonalizing the resulting quantum Hamiltonian. The computational details, although much more involved in this case, are essentially similar to that of the $LL$ model discussed in the previous section. Here one has also to 
employ the principal value prescription due to discontinuities in the wave-function and its derivatives. As explained in the previous section, we can implement the \emph{p.v.} prescription by using the operator product \eqref{sp:sklyanin_prod_general} instead. Therefore, we can reformulate one of the main results of \cite{Melikyan:2014mfa} in terms of the operator product \eqref{sp:sklyanin_prod_general} as follows: after the aforementioned field transformation, the quantum Hamiltonian which can be explicitly diagonalized has the form:
\begin{align}
\mathcal{H} & =-\frac{i}{2}\left(\psi_{i_{1}}^{\dagger}\gamma_{i_{1}i_{2}}^{3}\partial_{1}\psi_{i_{2}}-\partial_{1}\psi_{i_{1}}^{\dagger}\gamma_{i_{1}i_{2}}^{3}\psi_{i_{2}} \right)+m\psi_{i_{1}}^{\dagger}\gamma_{i_{1}i_{2}}^{0}\psi_{i_{2}}\nonumber \\
 & \qquad+\frac{i}{2}g_{2}\left(\psi_{i_{1}}^{\dagger}\psi_{j_{1}}^{\dagger}\gamma_{i_{1}i_{2}}^{0}\gamma_{j_{1}j_{2}}^{3}\psi_{i_{2}}\partial_{1}\psi_{j_{2}}-\psi_{i_{1}}^{\dagger}\partial_{1}\psi_{j_{1}}^{\dagger}\gamma_{i_{1}i_{2}}^{0}\gamma_{j_{1}j_{2}}^{3}\psi_{i_{2}}\psi_{j_{2}} \right)\nonumber \\
 & \qquad+\frac{g_{2}}{2m} \left(\psi_{i_{1}}^{\dagger}\psi_{j_{1}}^{\dagger}\gamma_{i_{1}i_{2}}^{0}\gamma_{j_{1}j_{2}}^{0}\partial_{1}\psi_{i_{2}}\partial_{1}\psi_{j_{2}}+\partial_{1}\psi_{i_{1}}^{\dagger}\partial_{1}\psi_{j_{1}}^{\dagger}\gamma_{i_{1}i_{2}}^{0}\gamma_{j_{1}j_{2}}^{0}\psi_{i_{2}}\psi_{j_{2}} \right)\nonumber \\
 & \qquad-\left(\frac{g_{3}+2g_{2}^{2}}{8m}\right) \left(\psi_{i_{1}}^{\dagger}\psi_{j_{1}}^{\dagger}\partial_{1}\psi_{k_{1}}^{\dagger}\gamma_{i_{1}i_{2}}^{0}\gamma_{j_{1}j_{2}}^{0}\gamma_{k_{1}k_{2}}^{0}\psi_{i_{2}}\psi_{j_{2}}\partial_{1}\psi_{k_{2}} \right)\nonumber \\
 & \qquad+i\frac{g_{2}^{2}}{8m^{2}}\left(\psi_{i_{1}}^{\dagger}\psi_{j_{1}}^{\dagger}\partial_{1}\psi_{k_{1}}^{\dagger}\gamma_{i_{1}i_{2}}^{0}\gamma_{j_{1}j_{2}}^{0}\gamma_{k_{1}k_{2}}^{3}\psi_{i_{2}}\psi_{j_{2}}\partial_{1}^{2}\psi_{k_{2}} -\psi_{i_{1}}^{\dagger}\psi_{j_{1}}^{\dagger}\partial_{1}^{2}\psi_{k_{1}}^{\dagger}\gamma_{i_{1}i_{2}}^{0}\gamma_{j_{1}j_{2}}^{0}\gamma_{k_{1}k_{2}}^{3}\psi_{i_{2}}\psi_{j_{2}}\partial_{1}\psi_{k_{2}} \right)\nonumber \\
 & \qquad-i\frac{g_{2}^{2}}{2m^{2}}\left(\psi_{i_{1}}^{\dagger}\psi_{j_{1}}^{\dagger}\partial_{1}\psi_{k_{1}}^{\dagger}\gamma_{i_{1}i_{2}}^{0}\gamma_{j_{1}j_{2}}^{0}\gamma_{k_{1}k_{2}}^{3}\psi_{i_{2}}\partial_{1}\psi_{j_{2}}\partial_{1}\psi_{k_{2}}-\psi_{i_{1}}^{\dagger}\partial_{1}\psi_{j_{1}}^{\dagger}\partial_{1}\psi_{k_{1}}^{\dagger}\gamma_{i_{1}i_{2}}^{0}\gamma_{j_{1}j_{2}}^{0}\gamma_{k_{1}k_{2}}^{3}\psi_{i_{2}}\psi_{j_{2}}\partial_{1}\psi_{k_{2}}\right)\nonumber \\
 & \qquad+\frac{g_{2}^{3}}{2m^{3}}\left(\psi_{i_{1}}^{\dagger}\psi_{j_{1}}^{\dagger}\partial_{1}\psi_{k_{1}}^{\dagger}\partial_{1}\psi_{l_{1}}^{\dagger}\gamma_{i_{1}i_{2}}^{0}\gamma_{j_{1}j_{2}}^{0}\gamma_{k_{1}k_{2}}^{0}\gamma_{l_{1}l_{2}}^{0}\psi_{i_{2}}\psi_{j_{2}}\partial_{1}\psi_{k_{2}}\partial_{1}\psi_{l_{2}}\right).\label{aaf:transformed_hamiltonian}
\end{align}
We stress that the fermionic fields in \eqref{aaf:transformed_hamiltonian} are the regularized fields as in \eqref{sp:field_distribution}, and the operator product is understood to be the $\circ$-product. We omitted here for simplicity the index $\mathscr{F}$ in fermionic fields, as well as the explicit $\circ$-product symbol in \eqref{aaf:transformed_hamiltonian}. 

Some comments are in order. Firstly, in the quantum mechanical picture, the wave-functions (c.f. \eqref{sp:fntilde}) as well as their derivatives are not, as we discussed, continuous functions and satisfy a rather involved relation. A consequence of this relation is the fact that the quantum mechanical Hamiltonian is a self-adjoint operator. Secondly, it was shown that in the process one recovers the correct $S$-matrix \eqref{aaf:s_matrix}, which was previously found only from the perturbative calculations from \eqref{aaf:action}. The direct diagonalization provides a non-perturbative confirmation of the $S$-matrix.

Although this initial step of constructing the quantum Hamiltonian reproduces all the known results, it is still not obvious how to derive such quantum charges using the methods of integrable systems. The $AAF$ model is a non-ultralocal model, and no standard procedure exists to construct the quantum Hamiltonian and other quantum charges. The result above shows however, that whatever the method, while working with continuous quantum systems, it should automatically produce the quantum Hamiltonian \eqref{aaf:transformed_hamiltonian}, i.e., it should already contain the $\circ$-product from the beginning on a more fundamental level. In other words, if there exists some generalizations of Yang-Baxter relations for non-ultralocal systems, such operator relations should already be written in terms of the $\circ$-product \eqref{sp:sklyanin_prod_general}, as well as in terms of the $\mathscr{F}$-regularized fields (distributions).

As a demonstration of this point of view and its consequences, we consider in the next section the simpler model of a free massive fermionic field, for which the Lax pair and the corresponding algebra can be readily obtained from the $AAF$ model. Although a free model, it is still a non-ultralocal model of the same order as $AAF$ with a rather involved Lax pair, on the example of which we show how to carry out the quantum calculations, without the technical complications of the full $AAF$ model. We show below the relation of the $\circ$-product with normal ordering for this case, and explain how to obtain the quantum Hamiltonian \eqref{aaf:transformed_hamiltonian} reduced to the free case. Another consequence will be explored in the subsequent section, where we make a connection of the $\circ$-product of operators with Maillet's symmetrization procedure in the classical case.

\section{Free massive fermion model}
\label{sec:free}

The massive free fermion model can be obtained as a consistent reduction of the $AAF$ model by setting the coupling constants  $g_{2}=0$ and $g_{3}=0$ in the action \eqref{aaf:action} and in the Lax pair for the $AAF$ model. The explicit formulas and computational details are given in \cite{Melikyan:2012kj} and \cite{Melikyan:2014yma}. Here we only list the necessary results.

\subsection{Classical integrability}
\label{sec:free_c}

We start from the Lax pair obtained as a result of this reduction:
\begin{align}
		 L^{\scriptscriptstyle{(\tau)}}(x;\lambda) &= \hat{\xi}_{0}^{\scriptscriptstyle{(\tau)}}(x;\lambda) \mathbb{1} + \hat{\xi}_{1}^{\scriptscriptstyle{(\tau)}}(x;\lambda) \sigma^{3} + \hat{\Lambda}^{\scriptscriptstyle{(-)}}_{\tau}(x;\lambda) \sigma^+ + \hat{\Lambda}^{\scriptscriptstyle{(+)}}_{\tau}(x;\lambda) \sigma^-, \label{free:Lax_pair_L0} \\
	 L^{\scriptscriptstyle{(\sigma)}}(x;\lambda) &= \hat{\xi}_{0}^{\scriptscriptstyle{(\sigma)}}(x;\lambda) \mathbb{1} + \hat{\xi}_{1}^{\scriptscriptstyle{(\sigma)}}(x;\lambda) \sigma^{3} + \hat{\Lambda}^{\scriptscriptstyle{(-)}}_{\sigma}(x;\lambda) \sigma^+ + \hat{\Lambda}^{\scriptscriptstyle{(+)}}_{\sigma}(x;\lambda) \sigma^- \label{free:Lax_pair_L1}.
\end{align}
The explicit form of the functions $\hat{\xi}_{j}^{\scriptscriptstyle{(\sigma,\tau)}}(x;\lambda)$, $j=0,1$, and $\hat{\Lambda}_{\sigma,\tau}^{\scriptscriptstyle{(\pm)}}(x;\lambda)$ is given in appendix \ref{app:free_app}.
The classical algebra of transition matrices \eqref{aaf:T_algebra_symm} has been given in \cite{Melikyan:2014yma} for the infinite line case, and has the same structure as that of the full $AAF$ model \eqref{aaf:lax_general}. We stress that even in the free fermion case, the coefficients $A(x,y;\lambda,\mu)$, $B(x,y;\lambda,\mu)$ and $C(x,y;\lambda,\mu)$ are non-vanishing and nonlinear functions of the fermionic fields. Nonetheless, the algebra for the reduced monodromy matrix:
\begin{align}
	T(\lambda) &= \lim_{\substack{x \to +\infty\\ y \to -\infty}} \left[ e^{-\left(\hat{\xi}_{1}^{\scriptscriptstyle{(\sigma)}}(\lambda)\sigma^{3}x \right)} T(x,y;\lambda)e^{\left( \hat{\xi}_{1}^{\scriptscriptstyle{(\sigma)}}(\lambda)\sigma^{3}y\right)} \right], \label{free_q:reduced_mon}
\end{align}
can be easily obtained:
\begin{align}
	 \left\{ T(\lambda) \stackrel{\otimes}{,} T(\mu) \right\}_{M} = {u}_{+}(\lambda,\mu) \: T(\lambda) \otimes T(\mu) - T(\lambda) \otimes T(\mu) \:{u}_{-}(\lambda,\mu). \label{free:reduced_mon_alg}
\end{align}
The matrices ${u}_{+}(\lambda,\mu)$ and ${u}_{-}(\lambda,\mu)$ in \eqref{free:reduced_mon_alg} have the following form:
 \begin{align}
 	u_{+}(\lambda,\mu) &= \left( \begin{array}{cccc}
 			-\text{p.v.}\:a(\lambda,\mu) & 0 & 0 & 0 \\
 		 0 &  -b(\lambda,\mu) & i \pi c(\lambda) \delta(\lambda -\mu) & 0 \\
 			0 &  -i \pi c(\lambda) \delta(\lambda -\mu) & b(\lambda,\mu) & 0 \\
 			0 & 0 & 0 & \text{p.v.}\:a(\lambda,\mu)
 			\end{array} \right), \label{free:u_plus}
 \end{align}
and
\begin{align}
	u_{-}(\lambda,\mu) &= \left( \begin{array}{cccc}
			-\text{p.v.}\:a(\lambda,\mu) & 0 & 0 & 0 \\
		 0 &  -b(\lambda,\mu) & -i \pi c(\lambda) \delta(\lambda -\mu) & 0 \\
			0 &  i \pi c(\lambda) \delta(\lambda -\mu) & b(\lambda,\mu) & 0 \\
			0 & 0 & 0 & \text{p.v.}\:a(\lambda,\mu)
			\end{array} \right),\label{actang:u_minus}
\end{align}
with
\begin{align}
	a(\lambda,\mu) &:= \frac{\coth({\lambda - \mu})\sinh^{2}({\lambda + \mu})}
	{2k}\label{free:a},\\
	b(\lambda,\mu) &:=\frac{\sinh(2\left(\lambda +\mu)\right)}{4k} \label{free:b},\\
	c(\lambda) &:=-\frac{1}{2k}\sinh^{2}(2\lambda). \label{free:c}
\end{align}
One can use these formulas and the standard methods of integrable models to obtain the action-angle variables as well as the classical conserved (local and non-local) quantities \cite{Melikyan:2014yma}.

\subsection{Quantum integrability}
\label{sec:free_q}
For the quantum case, in the absence of a Yang-Baxter-like relation, we have first to carefully define the quantum transition matrix. Therefore, we start from the standard classical definition of the transition matrix $T(x,y;\lambda)$ via the following differential equations:
\begin{align}
	\partial_x T(x,y;\lambda) &= L^{\scriptscriptstyle{(\sigma)}}(x;\lambda) T(x,y;\lambda), \label{free_q:diff_eq_Tx} \\
	\partial_y T(x,y;\lambda) &= - T(x,y;\lambda) L^{\scriptscriptstyle{(\sigma)}}(y;\lambda), \label{free_q:diff_eq_Ty}\\
\lim_{x\to y}T(x,y;\lambda) &= \mathds{1}, \label{free_q:T_x_x}
\end{align}
where the classical Lax operator $L^{\scriptscriptstyle{(\sigma)}}(x;\lambda)$ has the form:
\begin{align}
	L^{\scriptscriptstyle{(\sigma)}}(x;\lambda) &= \left( \begin{array}{cccc}
			\hat{\xi}_{0}^{\scriptscriptstyle{(\sigma)}}(x;\lambda)+\hat{\xi}_{1}^{\scriptscriptstyle{(\sigma)}}(\lambda) & \hat{\Lambda}^{\scriptscriptstyle{(-)}}_{\sigma}(x;\lambda)  \\
		 \hat{\Lambda}^{\scriptscriptstyle{(+)}}_{\sigma}(x;\lambda) & \hat{\xi}_{0}^{\scriptscriptstyle{(\sigma)}}(x;\lambda) - \hat{\xi}_{1}^{\scriptscriptstyle{(\sigma)}}(\lambda)
			\end{array} \right).\label{free_q:Lax_classical}
\end{align}
Equivalently, one can define the transition matrix $T(x,y;\lambda)$ via the corresponding integral equations:
\begin{align}
T(x,y;\lambda)& =E(x-y,\lambda )+\int_{y}^{x}T(x,z;\lambda)U^{(0)}(z,\lambda)E(z-y,\lambda)dz , \label{free_q:F1} \\
T(x,y;\lambda)& =E(x-y,\lambda )+\int_{y}^{x}E(x-z,\lambda)U^{(0)}(z,\lambda)T(z,y;\lambda)dz,  \label{free_q:F2}
\end{align}
where in our case:
\begin{align}
E(x,\lambda ) &= e^{\hat{\xi}_{1}^{\scriptscriptstyle{(\sigma)}}(\lambda)\sigma ^{3}x},\label{free_q:def1} \\
L^{\scriptscriptstyle{(\sigma)}}(x;\lambda) &=U^{(1)}(\lambda )+U^{(0)}(x,\lambda ), \label{free_q:def2}\\
U_{1}(\lambda )& =\hat{\xi}_{1}^{\scriptscriptstyle{(\sigma)}}(\lambda)\sigma ^{3}\label{free_q:def2}, \\
U^{(0)}(z,\lambda )&=\left(
\begin{array}{cc}
\hat{\xi}_{0}^{\scriptscriptstyle{(\sigma)}}(x,\lambda ) & \hat{\Lambda}^{\scriptscriptstyle{(-)}}_{\sigma}(x,\lambda ) \\
\hat{\Lambda}^{\scriptscriptstyle{(+)}}_{\sigma}(x,\lambda ) & \hat{\xi}_{0}^{\scriptscriptstyle{(\sigma)}}(x,\lambda )
\end{array}
\right). \label{free_q:def4}
\end{align}

Now we turn our attention to the quantum case. By considering the Lax operator  \eqref{free_q:Lax_classical} with the fields replaced by the corresponding quantum operators treated as operator valued distributions as in  \eqref{sp:field_distribution},\footnote{Here and below we omit for simplicity the index $\mathscr{F}$ in fermionic fields.} we use the integral equations above to define iteratively the quantum transition matrix in each order of iteration.  According to our prescription (see discussion in section \ref{sec:sp}), we regularize the operator products by employing $\circ$-product \eqref{sp:sklyanin_prod}. Thus, our quantum Lax operator will have the same form as in \eqref{free_q:Lax_classical}, but with $\hat{\xi}^{\scriptscriptstyle{(\sigma)}}_0(x;\lambda)$ regularized via Sklyanin's product. Namely, we take (see appendix \ref{app:free_app} for notations):
\begin{align}
	\hat{\xi}^{\scriptscriptstyle{(\sigma)}}_0(x) = \frac{1}{4J} \left[ - \chi_3(x)\circ \chi_1'(x) + \chi_4(x)\circ \chi_2'(x) - \chi_1(x)\circ \chi_3'(x) + \chi_2(x)\circ \chi_4'(x) \right].\label{free_q:ksi0again_SP}
\end{align}

Since we are dealing with the free fermion theory, we pause here to analyze Sklyanin's $\circ$-product in details. The standard formulas (for $x \neq y$) read:
\begin{align}
	\psi_{\alpha}(x)\bar{\psi}_{\beta}(y) =& \normord{\psi_{\alpha}(x)\bar{\psi}_{\beta}(y)} -\; iS^{+}_{\alpha \beta}(x-y), \label{free_q:norm_ordering_plus} \\
	\bar{\psi}_{\beta}(y) \psi_{\alpha}(x) =& \normord{\bar{\psi}_{\beta}(y) \psi_{\alpha}(x)} - \; iS^{-}_{\alpha \beta}(x-y). \label{free_q:norm_ordering_minus}
\end{align}
The singular behavior of the product of fields when $x \to y$ is entirely contained in the functions $S^{\pm}_{\alpha \beta}(x-y)$. Their explicit form is:
\begin{align}
	S^{+}_{\alpha \beta}(x) &= \frac{i}{2} \int \frac{dp_{1}}{2 \pi}\left( \frac{\slashed{p} + m}{\omega(p)}\right) e^{-i px}, \label{free_q:S+}\\
S^{-}_{\alpha \beta}(x) &= \frac{i}{2} \int \frac{dp_{1}}{2 \pi}\left( \frac{\slashed{p} - m}{\omega(p)}\right) e^{i px}, \label{free_q:S+}
\end{align}
where $\omega(p):= \sqrt{(p_{1})^{2} +m^{2}}$.
Since the normal ordering is free of singularities, we obtain from \eqref{free_q:norm_ordering_plus}, \eqref{free_q:norm_ordering_minus} and \eqref{sp:sklyanin_prod}:
\begin{align}
	\psi_{\alpha}(x) \circ \bar{\psi}_{\beta}(x) &= \normord{\psi_{\alpha}(x)\bar{\psi}_{\beta}(x)} + \Gamma_{\alpha \beta}^{+}(x), \label{free_q:norm_ordering_plus_SP} \\
	\bar{\psi}_{\beta}(x) \circ {\psi}_{\alpha}(x) &= \normord{\bar{\psi}_{\beta}(x) {\psi}_{\alpha}(x)} + {\Gamma}_{\alpha \beta}^{-}(x),\label{free_q:norm_ordering_minus_SP}
\end{align}
where the functions $\Gamma^{\pm}_{\alpha \beta}(x)$ take the form:
\begin{align}
	\Gamma_{\alpha \beta}^{\pm}(x) := \lim_{\frac{\Delta}{2} \to \epsilon} \frac{(-i)}{(\Delta - \epsilon)^{2}}\left[\int\limits_{x-\frac{\Delta}{2}+\epsilon}^{x+\frac{\Delta}{2}}du \int\limits_{x-\frac{\Delta}{2}}^{u-\epsilon}dv \,\, S^{\pm}_{\alpha \beta}(u-v) + \int\limits_{x-\frac{\Delta}{2}}^{x+\frac{\Delta}{2} - \epsilon}du \int\limits_{u+\epsilon}^{x+\frac{\Delta}{2}}dv\,\,S^{\pm}_{\alpha \beta}(u-v)\right]. \label{free_q:Gamma_pm}
\end{align}

Thus, in the free fermion case the difference between Sklyanin's product and the normal ordering is merely a function. Moreover, we are interested in the equal time relations, and in this case the relation:
\begin{align}
	S^{+}_{\alpha \beta}(x)\Big|_{x^{0}} + S^{-}_{\alpha \beta}(x)\Big|_{x^{0}} = i \gamma^{0}\delta(x^{1}) \label{free_q:relation}
\end{align}
implies that (for a fixed regularization parameter $\epsilon$):
\begin{align}
	\psi_{\alpha}(x) \circ \bar{\psi}_{\beta}(x) = - \bar{\psi}_{\beta}(x) \circ {\psi}_{\alpha}(x),\label{free_q:SP_normal_order}
\end{align}
which also follows directly from the definition \eqref{sp:sklyanin_prod}. Hence, in the free fermion case the $\circ$-product enjoys the same properties as the normal ordering, and it is, as commented in section \ref{sec:sp}, a point splitting type regularization that symmetrically takes into account all points around the singular region.

It is not enough to regularize only the Lax operator \eqref{free_q:Lax_classical}. The integral equations themselves must be regularized, due to the product of the operators at the same point under the integral in \eqref{free_q:F1} and \eqref{free_q:F2}. Thus, together with the regularization of the Lax operator \eqref{free_q:Lax_classical}, where now the formula for $\hat{\xi}^{\scriptscriptstyle{(\sigma)}}_0(x;\lambda)$ must be regularized as in \eqref{free_q:ksi0again_SP}, one has to replace the integral equations \eqref{free_q:F1} and \eqref{free_q:F2} by their non-singular versions:
\begin{align}
\hat{T}(x,y;\lambda)& =E(x-y,\lambda)+\int_{y}^{x} \left[ \hat{T}(x,z;\lambda) \circ U^{(0)}(z,\lambda) \right] E(z-y,\lambda) \: dz,  \label{free_q:F1_SP} \\
\hat{T}(x,y;\lambda)& =E(x-y,\lambda )+\int_{y}^{x}E(x-z,\lambda) \left[ U^{(0)}(z,\lambda) \circ \hat{T}(z,y;\lambda) \right] \: dz.  \label{free_q:F2_SP}
\end{align}
Since the above expressions are well-defined, one can now find the differential equation satisfied by the quantum transition matrix $\hat{T}(x,y,\lambda)$:
\begin{align}
	\partial_x \hat{T}(x,y;\lambda) &= {L}_{1}(x;\lambda) \circ \hat{T}(x,y;\lambda), \label{free_q:diff_eq_Tx_SP} \\
	\partial_y \hat{T}(x,y;\lambda) &= - \hat{T}(x,y;\lambda) \circ {L}_{1}(y;\lambda). \label{free_q:diff_eq_Ty_SP}
\end{align}
In order to check these relations one has to verify the Leibniz rule for Sklyanin's product. This is done in appendix \ref{app:leibnitz}.

Having defined the quantum transition matrix $\hat{T}(x,y,\lambda)$, we can now solve the integral equations \eqref{free_q:F1_SP} and \eqref{free_q:F2_SP} iteratively. Denoting:
\begin{align}
\hat{T}(x,y;\lambda)&:= \left( \begin{array}{cccc}
				\hat{t}_{1}(x,y;\lambda) & \hat{t}_{2}(x,y;\lambda) \\
			\hat{t}_{3}(x,y;\lambda) &  \hat{t}_{4}(x,y;\lambda)
\end{array}\right),\label{free:T_t1_t4}
\end{align}
we find from \eqref{free_q:F1_SP} and \eqref{free_q:F2_SP} that in the second iteration the components $\hat{t}_{i}$; $i=1, \ldots, 4$ have the form:
% \hat{\xi}_{0}^{\scriptscriptstyle{(\sigma)}}(z;\lambda)
% \hat{\xi}_{1}^{\scriptscriptstyle{(\sigma)}}
% \hat{\Lambda}^{\scriptscriptstyle{(-)}}_{\sigma}(u;\lambda )
\begin{align}
	e^{-\hat{\xi}_{1}^{\scriptscriptstyle{(\sigma)}}(\lambda)\left( x-y\right) }\hat{t}_{1}(x,y;\lambda) &= 1 + \int_y^x dz \; \hat{\xi}_{0}^{\scriptscriptstyle{(\sigma)}}(z;\lambda) + \int_y^x dz \; \int_z^x du\; \hat{\xi}_{0}^{\scriptscriptstyle{(\sigma)}}(u;\lambda) \circ \hat{\xi}_{0}^{\scriptscriptstyle{(\sigma)}}(z;\lambda) \nonumber \\
	&+ \int_y^x dz \; e^{2 \hat{\xi}_{1}^{\scriptscriptstyle{(\sigma)}}(\lambda) z} \int_z^x du \; e^{-2\hat{\xi}_{1}^{\scriptscriptstyle{(\sigma)}}(\lambda) u} \hat{\Lambda}^{\scriptscriptstyle{(-)}}_{\sigma}(u;\lambda ) \circ \hat{\Lambda}^{\scriptscriptstyle{(+)}}_{\sigma}(z;\lambda ), \label{free_q:t1_quantum} \displaybreak[3] \\
	e^{-\hat{\xi}_{1}^{\scriptscriptstyle{(\sigma)}}(\lambda)\left( x+y\right) }\hat{t}_{2}(x,y;\lambda) &= \int_y^x dz\; e^{-2\hat{\xi}_{1}^{\scriptscriptstyle{(\sigma)}}(\lambda)z} \hat{\Lambda}^{\scriptscriptstyle{(-)}}_{\sigma}(z;\lambda ) + \int_y^x dz\; e^{-2 \hat{\xi}_{1}^{\scriptscriptstyle{(\sigma)}}(\lambda) z} \int_z^x du\; \hat{\xi}_{0}^{\scriptscriptstyle{(\sigma)}}(u;\lambda) \circ \hat{\Lambda}^{\scriptscriptstyle{(-)}}_{\sigma}(z;\lambda ) \nonumber \\
	&+\int_y^x dz\; \int_z^x du\; e^{-2  \hat{\xi}_{1}^{\scriptscriptstyle{(\sigma)}}(\lambda)u} \hat{\Lambda}^{\scriptscriptstyle{(-)}}_{\sigma}(u;\lambda ) \circ  \hat{\xi}_{0}^{\scriptscriptstyle{(\sigma)}}(z;\lambda), \label{free_q:t2_quantum} \displaybreak[3] \\
	e^{ \hat{\xi}_{1}^{\scriptscriptstyle{(\sigma)}}(\lambda)(x+y)} \hat{t}_{3}(x,y;\lambda) &= \int_y^x dz\; e^{2 \hat{\xi}_{1}^{\scriptscriptstyle{(\sigma)}}(\lambda)z}  \hat{\Lambda}^{\scriptscriptstyle{(+)}}_{\sigma}(z;\lambda) + \int_y^x dz\; e^{2 \hat{\xi}_{1}^{\scriptscriptstyle{(\sigma)}} (\lambda)z} \int_z^x du\;  \hat{\xi}_{0}^{\scriptscriptstyle{(\sigma)}}(u;\lambda) \circ \hat{\Lambda}^{\scriptscriptstyle{(+)}}_{\sigma}(z;\lambda ) \nonumber\\
	&+\int_y^x dz\; \int_z^x du\; e^{2 \hat{\xi}_{1}^{\scriptscriptstyle{(\sigma)}}(\lambda)u}  \hat{\Lambda}^{\scriptscriptstyle{(+)}}_{\sigma}(u;\lambda )\circ \hat{\xi}_{0}^{\scriptscriptstyle{(\sigma)}}(z;\lambda),  \label{free_q:t3_quantum} \displaybreak[3] \\
	e^{\hat{\xi}_{1}^{\scriptscriptstyle{(\sigma)}}(\lambda)\left( x-y\right) }\hat{t}_{4}(x,y;\lambda) &= 1 + \int_y^x dz \; \hat{\xi}_{0}^{\scriptscriptstyle{(\sigma)}}(z;\lambda) + \int_y^x dz \; \int_z^x du\; \hat{\xi}_{0}^{\scriptscriptstyle{(\sigma)}}(u;\lambda) \circ \hat{\xi}_{0}^{\scriptscriptstyle{(\sigma)}}(z;\lambda) \nonumber \\
	&+ \int_y^x dz \; e^{-2 \hat{\xi}_{1}^{\scriptscriptstyle{(\sigma)}}(\lambda) z} \int_z^x du \; e^{2\hat{\xi}_{1}^{\scriptscriptstyle{(\sigma)}}(\lambda) u} \hat{\Lambda}^{\scriptscriptstyle{(+)}}_{\sigma}(u;\lambda ) \circ \hat{\Lambda}^{\scriptscriptstyle{(-)}}_{\sigma}(z;\lambda ). \label{free_q:t4_quantum}
\end{align}
These are well-defined expressions as long as the regularization parameter $\epsilon$ is a fixed number different from zero. Thus, one can easily continue this iterative process and obtain well-defined components of the quantum transition matrix for any iteration order.

Using that the fields vanish at infinity, i.e., $\chi_i(z) \xrightarrow{z \to \pm \infty} 0$, $i=1,\ldots,4$, one can readily verify that the component $\hat{t}_{1} (\lambda)$ is a conserved quantity, expansion of which has the form:
\begin{align}
	\hat{t}_{1} (\lambda) &= 1 - \frac{i}{2} \cosh(2\lambda) \int_{-\infty}^{+\infty} dz \; \mathcal{P}(z) + \frac{i}{2k}\sinh(2\lambda) \cosh(2\lambda) \int_{-\infty}^{+\infty} dz \; \mathcal{Q}(z) \nonumber \\
	&- \frac{i}{2k} \sinh(2\lambda) \int_{-\infty}^{+\infty} dz \; \mathcal{H}_0(z) + \sinh^2(2\lambda) N(\lambda) + O(\chi^4) \label{free:expansion}
\end{align}
and contains the quantum Hamiltonian $\mathcal{H}_{0}$ for free fermion model,
\begin{align}
\mathcal{H}_0(z) &= J -\frac{k}{2J} \left[\chi_2(z) \circ \chi_3'(z) -\chi_2'(z) \circ \chi_3(z) - \chi_1(z) \circ \chi_4'(z) +\chi_1'(z) \circ \chi_4(z)\right] \nonumber \\
	&+ \chi_2(z) \circ \chi_4(z) - \chi_1(z) \circ \chi_3(z). \label{free:free_fermion_hamiltonian}
\end{align}
The other terms in the expansion \eqref{free:expansion} correspond to the momentum and the charge operators,
\begin{align}
	\mathcal{P}(z) &= -\frac{i}{J} \left[ \chi_3(z) \circ \chi_1'(z) + \chi_4(z) \circ \chi_2'(z) \right], \label{free:free_fermion_momentum}\\
	\mathcal{Q}(z) &= - \chi_2(z) \circ \chi_4(z) - \chi_1(z) \circ \chi_3(z), \label{free:free_fermion_charge}
\end{align}
as well as conserved non-local charges,
\begin{align}
	N(\lambda) &= \frac{1}{2k} \int_{-\infty}^{+\infty} dz \; \left[ \chi_1(z) \circ \chi_4(z)-\chi_2(z) \circ \chi_3(z) \right] \nonumber \\
	&+ \frac{J}{k^2} \int_{-\infty}^{+\infty} dz \; e^{2 \hat{\xi}_{1}^{\scriptscriptstyle{(\sigma)}}(\lambda)z} \int_{z}^{+\infty} du \; e^{-2 \hat{\xi}_{1}^{\scriptscriptstyle{(\sigma)}}(\lambda)u} \left[ l_3^2(\lambda)  \chi_2(u) \circ \chi_4(z) + l_4^2(\lambda)  \chi_1(u) \circ \chi_3(z) \right. \nonumber \\
	&+ \left. il_3(\lambda) l_4(\lambda) \left(  \chi_1(u) \circ \chi_4(z) -  \chi_2(u) \circ \chi_3(z) \right) \right]. \label{free:free_fermion_non_local_charges}
\end{align}
The expression \eqref{free:free_fermion_hamiltonian} coincides with the free part of the quantum Hamiltonian for the $AAF$ model \eqref{aaf:transformed_hamiltonian}, with operator product already regularized via Sklyanin's product
\begin{align}
\mathcal{H}_{0}  =-\frac{i}{2}\left(\psi_{i_{1}}^{\dagger}\gamma_{i_{1}i_{2}}^{3} \circ \partial_{1}\psi_{i_{2}}-\partial_{1}\psi_{i_{1}}^{\dagger}\gamma_{i_{1}i_{2}}^{3} \circ \psi_{i_{2}} \right)+m\psi_{i_{1}}^{\dagger}\gamma_{i_{1}i_{2}}^{0} \circ \psi_{i_{2}}.\label{free:free_fermion_hamiltonian_transformed}
\end{align}
Thus, we have shown, on the example of the free fermion model how to obtain the quantum conserved charges, and in particular, the quantum Hamiltonian, from the quantum transition matrix with products regularized via Sklyanin product. 

We conclude this section with a remark concerning the interacting case. One can, in principle, repeat the above calculations for the interacting theory, i.e., the full $AAF$ model. Starting from the integral equations \eqref{free_q:F1_SP} and \eqref{free_q:F2_SP}, the quantum transition matrix is a well-defined object from the beginning, and one does not have to worry about singular expressions. For interacting fields, however, we do not have a simple relation between Sklyanin's product and the normal ordering, as it is the case for the free fields, cf. \eqref{free_q:norm_ordering_plus_SP} and \eqref{free_q:norm_ordering_minus_SP}. In this case, the normal ordering does not solve the singularity problem, and one has to use the \emph{normal product} introduced by Zimmermann (for a review see \cite{Zimmermann:1970is}) instead. The point splitting regularization in the interacting theory can be expressed via the operator product expansion \cite{Wilson:1972ee,Zimmermann:1970is,Wilson:1970pq,Lowenstein:1970jh} as:
\begin{align}
	\psi(x) \bar{\psi}(y) = \sum\limits_{n=1}^{\infty} C_{n}(x-y) \mathcal{O}(x),\label{free_q:ope}
\end{align}
where $\mathcal{O}(x)$ correspond to composite local operators defined via normal product, and the singularities are exhibited in the functions $C_{n}(x-y)$ when taking the limit $x \to y$.

Some comments are in order. First, the expansion in \eqref{free_q:ope} can be strictly proven for (BPHZ) renormalizable theories in each order of the perturbation theory. Although there is no strict proof of the renormalizability of the $AAF$ model, it has been proposed to be such a theory in \cite{Alday:2005jm}, and the recently obtained explicit diagonalization of the quantum Hamiltonian \eqref{aaf:transformed_hamiltonian} makes this proposal more plausible. Then proceeding exactly as in the free case, by smearing the product $\psi(x) \bar{\psi}(y)$ over the entire region around the singular strip, we can write:
\begin{align}
	\psi(x) \circ \bar{\psi}(x) = \sum\limits_{n=1}^{\infty} \xi_n({\epsilon}) \mathcal{O}(x),\label{free_q:ope2}
\end{align}
where $\epsilon$ is the regularization parameter entering into the definition of $\circ$-product \eqref{sp:sklyanin_prod}, and $\xi_n({\epsilon})$ are functions obtained from $C_{n}(x-y)$, which diverge in the limit $\epsilon \to 0$. Thus, in perturbation theory, one can naturally relate Sklyanin's product to renormalized composite operators $\mathcal{O}(x)$, which can be written in terms of Zimmermann's normal products and found perturbatively in each order. It would be interesting to carry out such explicit analysis for integrable models, as in this case the diagrammatic analysis significantly simplifies due to powerful non-renormalization theorems (see for details \cite{Klose:2006dd,Thacker:1974kv,Thacker:1976vp,Melikyan:2008ab,Melikyan:2011uf}).

\section{Quantum algebra of transition matrices}\label{sec:qa}

We are finally in a position to conjecture the form of the quantum algebra of transition matrices for non-ultralocal systems directly in the continuous case. As we discussed previously, one of the main difficulties is to formulate a regularization which consistently reduces to the classical algebra \eqref{aaf:T_algebra_symm} in the classical limit. Indeed, it is not obvious how a commutator between two operators, or in general $n$-nested commutators, restores Maillet's symmetrization prescription  \eqref{aaf:nested_brackets} and \eqref{aaf:symmetrization} upon taking the classical limit. Bellow, we consider the classical limit of a quantum commutator regularized in terms of Sklyanin's product and its relation to the Maillet bracket to propose the form of the quantum algebra of transition matrices to models such as the $AAF$ model or the free fermion model.

To obtain the aforementioned relation, we consider here, following \cite{Freidel:1991jx,Freidel:1991jv}, the simpler case of the general classical algebra \eqref{aaf:T_algebra_symm}:
\begin{align}
		\{T_{1}(x,y;\lambda), \; T_{2}(x,y;\mu) \}_{{M}}
		&= a_{12}(\lambda,\mu) \; T_{1}(x,y;\lambda)T_{2}(x,y;\mu)
		- T_{1}(x,y;\lambda)T_{2}(x,y;\mu) \;d_{12}(\lambda,\mu), \notag \\
		\{T_{1}(x,y;\lambda), \;T_{2}(y,z;\mu) \}_{{M}} &= T_{1}(x,y;\lambda) \; b_{12}(\lambda,\mu) \; T_{2}(y,z;\mu), \label{sp:Maillet_classical_algebra}
\end{align}
and recall the lattice algebra for the corresponding quantum case proposed by Freidel and Maillet in \cite{Freidel:1991jx,Freidel:1991jv}. It has the following form:
\begin{align}
\hat{A}_{12}T^{(n)}_{1}T^{(n)}_{2} &=T^{(n)}_{2} T^{(n)}_{1}\hat{D}_{12}, \label{qa:lat1}\\
T^{(n)}_{1}T^{(n+1)}_{2} &= T^{(n+1)}_{2} \hat{C}_{12}T^{(n)}_{1},\label{qa:lat2}\\
\left[ T^{(n)}_{1} ,T^{(m)}_{2} \right] &= 0, \quad \text{for  } |n-m| > 1. \label{qa:lat3}
\end{align}
Using the quasi-classical expansions
\begin{align*}
\hat{A}_{12}=1+ i \hbar a_{12}+ \ldots,
\end{align*}
 for all matrices in \eqref{qa:lat1}-\eqref{qa:lat3}, one obtains the classical lattice algebra:
\begin{align}
	\left\{ T_1^{(n)}(\lambda)  , T_2^{(n)}(\mu) \right\} &=  a(\lambda,\mu) \: T_1^{(n)}(\lambda)  T_2^{(n)}(\mu) - T_1^{(n)}(\lambda)  T_2^{(n)}{}(\mu) \: d(\lambda,\mu), \label{qa:lat1_cl}  \\
	\left\{ T_1^{(n)}(\lambda) , T_2^{(n+1)}(\mu) \right\} &=  - T_2^{(n+1)}(\mu) \: c(\lambda,\mu) \:   T_1^{(n)}(\lambda)  \label{qa:lat2_cl}\\
	\left\{ T_1^{(n)}(\lambda) , T_2^{(m)}(\mu) \right\} &= 0, \quad \text{for  } |n-m| > 1, \label{qa:lat3_cl}
\end{align}
where $T^{(n)}(\lambda) \equiv T(x_{n+1}, x_{n}; \lambda)$ is defined so as to make a connection with the continuous algebra \eqref{sp:Maillet_classical_algebra}. In passing from the quantum algebra \eqref{qa:lat1}-\eqref{qa:lat3} to the classical one \eqref{qa:lat1_cl}-\eqref{qa:lat3_cl}, one clearly obtains the usual Poisson brackets, which are not symmetrized according to Maillet's prescription \eqref{aaf:PB_sym}, as the lattice spacing already regulates all products. Moreover, the Jacobi identity for the classical lattice algebra \eqref{qa:lat1_cl}-\eqref{qa:lat3_cl} can be understood as a consequence of the consistency conditions (Yang-Baxter relations) satisfied by the quantum algebra \eqref{qa:lat1}-\eqref{qa:lat3} \cite{Freidel:1991jx,Freidel:1991jv}. Nonetheless, there remains the problem of restoring the symmetrization prescription upon removing the lattice regularization.

To solve this problem, we use the results explained in the previous sections to bypass the lattice reformulation of the quantum theory in favor of regularizing the continuous theory with Sklyanin's product \eqref{sp:sklyanin_prod} and \eqref{sp:sklyanin_prod_general}. The main idea is to formulate the quantum algebra of transition matrices $T(x,y;\lambda)$ so that the singular products are replaced by well-defined $\circ$-products. 

First, we make a key observation and show that the commutator of two operator-valued functions regularized with Sklyanin's product goes to the symmetrized Poisson brackets \eqref{aaf:PB_sym} in the classical limit. In this way, Maillet's symmetrization prescription appears naturally in the classical continuous theory from simply regularizing the singular operator products in the quantum case. Indeed, writing explicitly the commutator between two operator-valued functions $\hat{A}(x)$ and $\hat{B}(x)$ in terms of the definition \eqref{sp:sklyanin_prod}, we obtain:
\begin{equation}
\left[\hat{A}(x)\stackrel{\circ}{,}  \hat{B}(x)\right]
=  \lim_{\nicefrac{\Delta}{2} \to \epsilon} \frac{1}{(\Delta - \epsilon)^{2}}\left( \iint\limits_{\Delta S_{1}}d\zeta d\xi \left[ \hat{A}(\zeta), \hat{B}(\xi) \right] + \iint\limits_{\Delta S_{2}}d\zeta d\xi \left[ \hat{A}(\zeta), \hat{B}(\xi) \right] \right). \label{qa:sklyanin_prod_commutator}
\end{equation}
In the classical limit, denoted by CL bellow, equation \eqref{qa:sklyanin_prod_commutator} becomes:
\begin{equation}
\left[\hat{A}(x)\stackrel{\circ}{,}  \hat{B}(x)\right]_{\substack{CL}}
= \lim_{\nicefrac{\Delta}{2} \to \epsilon} \frac{1}{(\Delta - \epsilon)^{2}}\left( \iint\limits_{\Delta S_{1}}d\zeta d\xi \left\{{A}(\zeta), {B}(\xi) \right\} + \iint\limits_{\Delta S_{2}}d\zeta d\xi \left\{ {A}(\zeta), {B}(\xi) \right\} \right), \label{qa:sklyanin_prod_PB}
\end{equation}
where $A(\zeta)$ and $B(\xi)$ are already the corresponding classical functions, and $\left\{ A(\zeta), B(\xi) \right\}$ in the right-hand side is the usual Poisson bracket. It is then clear, by invoking the mean value theorem in the regions $\Delta S_{1}$ and $\Delta S_{2}$, where the integrands are smooth functions, that in the limit $\Delta \to \frac{\epsilon}{2}$, followed by the limit $\epsilon \to 0$, the right-hand side of \eqref{qa:sklyanin_prod_PB} reduces to:
\begin{align}
	\left[\hat{A}(x)\stackrel{\circ}{,}  \hat{B}(x)\right]_{\substack{CL}}
=  \frac{1}{2}  \lim_{\epsilon \to 0} \left( \left\{ A(x+\epsilon), B(x - \epsilon) \right\} + \left\{ A(x-\epsilon), B(x+\epsilon) \right\} \right).  \label{qa:SP_symmetrization}
\end{align}
Finally, comparing this formula with Maillet's definition of the symmetrized Poisson bracket \eqref{aaf:PB_sym} we conclude that:
\begin{align}
	\left[\hat{A}(x)\stackrel{\circ}{,}  \hat{B}(x)\right]_{\substack{CL}} = \left\{ A(x),B(x) \right\}_{M}. \label{qa:SP_sym_connection}
\end{align}

This simple observation shows the connection between Sklyanin's product in the quantum case, and Maillet's \emph{ad hoc} construction of symmetrized Poisson brackets. Namely, the classical limit of the commutator, regularized via Sklyanin's product, reproduces precisely the symmetrized Poisson bracket. Of course, if the integrable system is ultralocal, then both terms in \eqref{qa:SP_sym_connection} coincide, and Maillet's symmetrization procedure reduces to the usual Poisson brackets. This consideration can also be easily generalized, and the $n$-nested Poisson brackets \eqref{aaf:nested_brackets} can be similarly obtained from the general Sklyanin product \eqref{sp:sklyanin_prod_general}.

To conclude this section, we speculate on the form of the quantum continuous algebra. Namely, we propose the following quantum algebra, naturally reproducing the classical Maillet algebra \eqref{sp:Maillet_classical_algebra}:
\begin{align}
\hat{A}_{12}T_{1}(x,y;\lambda) \; {\circ} \;T_{2}(x,y;\mu) &=T_{2}(x,y;\mu) \; {\circ} \; T_{1}(x,y;\lambda)\hat{D}_{12}, \label{qa:nolat1_sp}\\
T_{1}(y,z;\lambda) \; {\circ} \; T_{2}(x,y;\mu) &= T_{2}(x,y;\mu) \; {\circ} \; \hat{C}_{12}T_{1}(y,z;\lambda).\label{qa:nolat2_sp}
\end{align}
These relations are well defined due to Sklyanin's product, and upon using as before the quasi-classical expansion for the matrices $(A,B,C,D)$, one obtains:
\begin{align}
	\left[ T_1(x,y;\lambda) \stackrel{\circ}{,} T_2(x,y;\mu) \right] & = -i a_{12}(\lambda,\mu) \; T_1(x,y;\lambda) \circ T_2(x,y;\mu) \label{qa:equal_interval} \\ &+ i T_2(x,y;\mu) \circ T_1(x,y;\lambda)\; d_{12}(\lambda,\mu) ,\notag \\
	\left[ T_1(y,z;\lambda)\stackrel{\circ}{,}  T_2(x,y;\mu) \right] &= i T_2(x,y;\mu)\circ c_{12}(\lambda,\mu) \; T_1(y,z;\lambda) \label{qa:adjacent_interval}.
\end{align}
Then, invoking the relation between Maillet's symmetrized brackets and the quantum commutator regularized by means of Sklyanin's product \eqref{qa:SP_sym_connection}, we can accordingly conclude that, in the classical limit, the quantum algebra given by \eqref{qa:equal_interval}  and \eqref{qa:adjacent_interval} reduces to the classical algebra \eqref{sp:Maillet_classical_algebra}. Moreover, the consistency conditions for the classical algebra \eqref{sp:Maillet_classical_algebra} derived by Freidel and Maillet in \cite{Freidel:1991jx,Freidel:1991jv} from the classical Jacobi identity follow similarly from the corresponding quantum Jacobi identity. We refer the reader to appendix \ref{app:ji} for a detailed discussion of the Jacobi identity for the quantum algebra  \eqref{qa:equal_interval}  and \eqref{qa:adjacent_interval}.

%Thus, we have shown how to obtain the classical algebra \eqref{sp:Maillet_classical_algebra} in terms of Maillet's symmetrized Poisson brackets from the quantum relations \eqref{qa:equal_interval} and \eqref{qa:adjacent_interval} involving commutators with respect to Sklyanin's product. This result allows us to interpret Maillet's symmetrisation prescription for Poisson brackets \eqref{intro:nested_brackets}, \eqref{intro:symmetrization} simply as the consequence of the regularisation of the singular operator product at the same point in the quantum case, when taking the classical limit.

\section{Conclusion}\label{sec:conclusion}

In this paper we have considered the quantization problem of continuous non-ultralocal integrable models, such as the $AAF$ model, without resorting to any lattice discretization. To do so, we have shown that it is necessary to treat the quantum fields as distributions, and employ a regularized product of operators - Sklyanin's product, in order to avoid singularities in diagonalization procedure, as well as to correctly reproduce the $S$-matrix known from perturbative calculations. Sklyanin's product corresponds essentially to a ``smeared'' product of operator-valued functions over a small neighborhood around the singular region, and it is naturally related, as discussed in the text, with renormalized local composite operators, defined via Zimmermann's normal products. As an example of this procedure, we were able to extract the quantum Hamiltonian as well as other conserved charges for the free fermion model directly from the quantum trace identities. In particular, the quantum Hamiltonian thus obtained coincides with previous results \cite{Melikyan:2014mfa}. In addition, we  demonstrated that the  quantum algebra of transition matrices, written in terms of the $\circ$-product consistently reproduces Maillet's symmetrization prescription for non-ultralocal integrable systems in the classical limit.

We outline here the possible future directions. As it is well known, the key obstacle in formulating a well-defined quantum algebra for transition matrices, corresponding to the classical expressions \eqref{sp:Maillet_classical_algebra}, is Schwartz's theorem on the impossibility of defining a product of two distributions with natural properties.\footnote{For an overview, see, for example, \cite{Zeidler:2006rw,Zeidler:2009zz}.} Although it is possible to define a product of distributions in some exceptional cases, for instance when their singular supports are disjoint, in general one must look for extensions of Schwartz's distribution theory. Out of several different approaches to evade Schwartz's impossibility theorem, two look specially promising: the microlocal analysis based on the concept of wavefront sets (for a review see \cite{brouder:2014jp} and references therein) and Colombeau algebras \cite{colombeau1990,colombeau2000new}. The former has been used in the context of quantum field theory, making it possible to define the product of distributions. However, even in this approach some interesting products of distributions, e.g., the powers of the Dirac delta distribution, remain ill-defined.  As for the latter, the space of Schwartz distributions is embedded into an associative algebra which satisfies the Leibniz's rule, nevertheless the association between a distribution and an element of the Colombeau algebra is not always unique. Thus, it is still an open and interesting question whether any of these approaches will prove to be successful in the context of the quantization of continuous non-ultralocal integrable systems. We also mention here that another compelling possibility to approach this problem lies within Sato's theory of hyperfunctions \cite{sato:1959av,sato:1969bb}, which also contains Schwartz's distribution theory and relies on the boundary behavior of analytic functions. 

Furthermore, as we discussed earlier at the end of section \ref{sec:sp}, Sklyanin's product is essentially a point splitting procedure taken uniformly over the entire region around the singularity strip. Thus, instead of taking the limit $x \to y$ symmetrically at the end of the calculation, as it is in the standard case of the point splitting procedure for the axial fermionic current \eqref{sp:anomaly}, we simply take into account all the points around the singularity equally from the beginning. In principle, for a renormalizable theory the quantum field equations can be written in terms of the Zimmermann's normal product in all orders. It is, therefore, possible to define the quantum Lax operator and the quantum transition matrix only in terms of these normal products. However, this is valid only for $BPHZ$ renormalizable theories, and for models such as the $AAF$ model, there is no proof of such renormalizability. Nevertheless, it is an interesting problem to formulate quantum integrable models (the Lax operator, the transition matrices, etc.) in terms of only Zimmermann's normal products. In general, in the absence of renormalizability of the theory, we can only use the generic Sklyanin's product, which simply homogeneously removes the singular region. Nonetheless, it is not immediately clear if Sklyanin's product is associative, presenting therefore an interesting problem to consider in the context of Schwarz's impossibility theorem for defining an associative product of distributions.\footnote{For a recent discussion of the problem of associativity in the context of operator product expansion, see \cite{Holland:2015tia}.} 

Finally, a related issue that will be interesting to consider regards the ultralocalization of second order non-ultralocal algebras such as the ones appearing in the $AAF$ model. Even though, a purely classical problem, it is the first step in the usual approach to non-ultralocal models, and it should, therefore, provide also a good insight in the quantization of second order non-ultralocal models. The natural starting point should be to consider a generalization of either the generalized $FR$ ultralocalization technique identified by \cite{Delduc:2012qb,Delduc:2012vq} in the context of sigma models or the procedure proposed by \cite{Alekseev:1991wq} for the $WZNW$ model to accommodate for the second derivative of the delta function appearing in \eqref{aaf:lax_general}. Moreover, it may also help us understand why the algebra of transition matrices has the same structure for both first and second order non-ultralocal systems. We hope to report progress on this direction soon.

\section*{Acknowledgments} G.W. would like to thank D. Guariento for useful discussions. The work of A.M. is partially supported by CAPES.

\section*{Appendices} \addcontentsline{toc}{section}{Appendices}
\appendix

\section{Notations}\label{app:free_app}

In this paper we consider results derived from two different but equivalent descriptions of the $AAF$ model. The goal of this appendix is, therefore, to fix our notations and show how these two descriptions are related. The action \eqref{aaf:action}
\begin{align}
	 S &= \int dy^0 \: \int _0^J dy^1 \: \left[ i \bar{\psi} \delslash \psi \: - m \bar{\psi} \psi + \frac{g_2}{4m} \epsilon^{\alpha \beta} \left( \bar{\psi}
	\partial_{\alpha} \psi \; \bar{\psi}\: \gamma^3
	\partial_{\beta} \psi -
	\partial_{\alpha}\bar{\psi} \psi \;
	\partial_{\beta} \bar{\psi}\: \gamma^3 \psi \right) \right.- \nonumber \\
	&- \left. \frac{g_3}{16m} \epsilon^{\alpha \beta} \left(\bar{\psi}\psi\right)^2
	\partial_{\alpha}\bar{\psi}\:\gamma^3
	\partial_{\beta}\psi \right], \label{free_app:action_pert}
\end{align}
was originally used in the papers \cite{Klose:2006dd,Melikyan:2011uf} to study the $AAF$ model from the perspective of perturbative quantum field theory. In \eqref{free_app:action_pert} $g_2$ and $g_3$ stand for the coupling constants of the model and the mass $m$ is related to the 't Hooft coupling $\lambda'$ via:
\begin{align}
	m = \frac{2 \pi}{\sqrt{\lambda'}}.
\end{align}
Here, the following representation for the Dirac matrices is employed
\begin{align}
	\label{free_app:gamma_matrices} \gamma^0 = \left(
	\begin{array}{cc}
		0 & 1 \\
		1 & 0
	\end{array}
	\right), \quad \gamma^1 = \left(
	\begin{array}{cc}
		0 & -1 \\
		1 & 0
	\end{array}
	\right), \quad \gamma^3 = \gamma^0 \gamma^1.
\end{align}
The action \eqref{free_app:action_pert} can be deduced from the original lagrangian proposed in \cite{Alday:2005jm}
\begin{align}
	\label{free_app:aaf_lagrangian_original} \mathscr{L} &= -J - \frac{iJ}{2} \left(\bar{\chi} \rho^0
	\partial_0 \chi -
	\partial_0 \bar{\chi} \rho^0 \chi \right) + i \kappa \left(\bar{\chi}\rho^1
	\partial_1 \chi -
	\partial_1 \bar{\chi} \rho^1 \chi \right) + J\bar{\chi}\chi  \nonumber \\
	&+ \frac{\kappa g_{2}}{2} \epsilon^{\alpha \beta} \left( \bar{\chi}
	\partial_{\alpha} \chi \; \bar{\chi} \rho^5
	\partial_{\beta} \chi -
	\partial_{\alpha}\bar{\chi} \chi \;
	\partial_{\beta} \bar{\chi} \rho^5 \chi \right) - \frac{\kappa g_{3}}{8} \epsilon^{\alpha \beta} \left(\bar{\chi}\chi\right)^2
	\partial_{\alpha}\bar{\chi}\rho^5
	\partial_{\beta}\chi,
\end{align}
where
%by performing some convenient transformation of the worldsheet coordinates and field redefinitions, which amount to
\begin{equation}\label{app:psi_chi_fields}
	\chi_1 = \frac{\psi_1 - \psi_2}{\sqrt{2}} \;, \quad \chi_2 = i \frac{\psi_1+ \psi_2}{\sqrt{2}},
\end{equation} 
and the representation of the Dirac matrices is
\begin{align}
	\label{free_app:rho_matrices} \rho^0 = \left(
	\begin{array}{cc}
		-1 & 0 \\
		0 & 1
	\end{array}
	\right), \quad \rho^1 = \left(
	\begin{array}{cc}
		0 & i \\
		i & 0
	\end{array}
	\right), \quad \rho^5 = \rho^0 \rho^1
\end{align}
with $\kappa = \frac{\sqrt{\lambda'}}{2}$. 
%to \eqref{free_app:gamma_matrices} and fixing $\kappa = \frac{\sqrt{\lambda'}}{2}$. 

The constant $J$ in \eqref{free_app:aaf_lagrangian_original} corresponds to the total angular momentum of the string in $S^5$. More importantly, the two lagrangians \eqref{free_app:action_pert} and \eqref{free_app:aaf_lagrangian_original} differ by some overall minus sign that was introduced in \cite{Melikyan:2011uf} in order to ensure a positive definite mode expansion. For all the details and a thorough derivation of \eqref{free_app:action_pert} from \eqref{free_app:aaf_lagrangian_original}, we refer the reader to the original papers \cite{Melikyan:2011uf,Alday:2005jm}.

The equations of motion for the free massive fermion following from \eqref{free_app:aaf_lagrangian_original} with $g_2=g_3=0$ can be described by the Lax pair \eqref{free:Lax_pair_L0} and \eqref{free:Lax_pair_L1}
\begin{align}
		 L^{\scriptscriptstyle{(\tau)}}(x;\mu) &= \hat{\xi}_{0}^{\scriptscriptstyle{(\tau)}}(x;\mu) \mathbb{1} + \hat{\xi}_{1}^{\scriptscriptstyle{(\tau)}}(x;\mu) \sigma^{3} + \hat{\Lambda}^{\scriptscriptstyle{(-)}}_{\tau}(x;\mu) \sigma^+ + \hat{\Lambda}^{\scriptscriptstyle{(+)}}_{\tau}(x;\mu) \sigma^-, \label{free_app:Lax_pair_L0} \\
	 L^{\scriptscriptstyle{(\sigma)}}(x;\mu) &= \hat{\xi}_{0}^{\scriptscriptstyle{(\sigma)}}(x;\mu) \mathbb{1} + \hat{\xi}_{1}^{\scriptscriptstyle{(\sigma)}}(x;\mu) \sigma^{3} + \hat{\Lambda}^{\scriptscriptstyle{(-)}}_{\sigma}(x;\mu) \sigma^+ + \hat{\Lambda}^{\scriptscriptstyle{(+)}}_{\sigma}(x;\mu) \sigma^- \label{free_app:Lax_pair_L1},
\end{align}
where the functions $\hat{\xi}_{j}^{\scriptscriptstyle{(\sigma,\tau)}}(x;\mu)$, $j=0,1$, and $\hat{\Lambda}_{\sigma,\tau}^{\scriptscriptstyle{(\pm)}}(x;\mu)$ have the following explicit form:
\begin{align}
	\hat{\xi}^{\scriptscriptstyle{(\sigma)}}_0 &= \frac{1}{4J} \left[ - \chi_3\chi_1' + \chi_4\chi_2' - \chi_1\chi_3' + \chi_2\chi_4' \right], \label{free_app:xis0}  \\
	\hat{\xi}^{\scriptscriptstyle{(\sigma)}}_1 &= \frac{il_2 J}{2 k}, \label{free:xis1} \\
\hat{\Lambda}^{\scriptscriptstyle{(-)}}_{\sigma} &= \frac{1}{\sqrt{J}} \left[ -l_{3}\chi_2'- il_{4}\chi_1' \right], \label{free_app:Lambdasm}  \\	
	\hat{\Lambda}^{\scriptscriptstyle{(+)}}_{\sigma} &= \frac{1}{\sqrt{J}} \left[ -l_{3}\chi_4'+ il_{4}\chi_3' \right],\label{free_app:Lambdasp}
\end{align}
and:
\begin{align}
	\hat{\xi}^{\scriptscriptstyle{(\tau)}}_0 &= \frac{i}{2J} \left[\chi_{3}\chi_{1}+\chi_{4}\chi_{2} \right]+ \frac{1}{4J}\left[-\chi_{3}\dot{\chi}_{1}-\chi_{1}\dot{\chi}_{3} +\chi_{4}\dot{\chi}_{2}  +\chi_{2}\dot{\chi}_{4} \right],  \label{free_app:ksi0tau}\\
	\hat{\xi}^{\scriptscriptstyle{(\tau)}}_1 &= -\frac{i l_{1}}{2}, \label{free_app:ksi1tau}\\
	\hat{\Lambda}^{\scriptscriptstyle{(-)}}_{\tau} &= -\frac{i}{\sqrt{J}}\left[l_{3}\chi_{2} - il_{4}\chi_{1} \right] - \frac{1}{\sqrt{J}} \left[l_{3}\dot{\chi}_{2} + il_{4}\dot{\chi}_{1}  \right], \label{free_app:Lambdataum} \\
	\hat{\Lambda}^{\scriptscriptstyle{(+)}}_{\tau} &= \frac{i}{\sqrt{J}}\left[l_{3}\chi_{4} + il_{4}\chi_{3} \right] - \frac{1}{\sqrt{J}} \left[l_{3} \dot{\chi}_{4} - il_{4}\dot{\chi}_{3}  \right].\label{free_app:Lambdataup}
\end{align}
Here we have denoted:
\begin{align}
	\chi = \left( \begin{array}{c}
		\chi_1\\
		\chi_2
	\end{array} \right), \quad \chi_{3}\equiv\chi_{1}^{*}, \quad \chi_{4}\equiv\chi_{2}^{*}.\label{free_app:notations_chis}
\end{align}
The dependence on the spectral parameter $\mu$ is encoded in the functions $l_{i}$ \cite{Alday:2005jm,Alday:2005gi,Arutyunov:2009ga}:
\begin{align}\label{free_app:l_i}
	l_0 = 1, \quad l_1 = \cosh(2\mu), \quad l_2 = -\sinh(2\mu), \quad l_3 = \cosh (\mu), \quad l_4 = \sinh(\mu).
\end{align}
The constant $k=\sqrt{\lambda^{'}}$.

\section{Leibnitz rule for Sklyanin's product}
\label{app:leibnitz}

In this appendix we show that the Leibnitz rule is valid for Sklyanin's product. Namely, for two operator-valued functions $A(x)$ and $B(x)$, one has:
\begin{align}
	\partial_{x} \left[A(x) \circ B(x) \right] = \partial_{x} \left[A(x) \right] \circ B(x) +  A(x) \circ \partial_{x} \left[ B(x) \right]. \label{leibnitz:rule}
\end{align}
To prove this formula, we write Sklyanin's product \eqref{sp:sklyanin_prod} in the following form:
\begin{align}
	A(x) \circ B(x) = \lim_{\frac{\Delta}{2} \to \epsilon} \frac{1}{(\Delta - \epsilon)^{2}}\left[\int\limits_{x-\frac{\Delta}{2}+\epsilon}^{x+\frac{\Delta}{2}}du\;A(u) \int\limits_{x-\frac{\Delta}{2}}^{u-\epsilon}dv \; B(v) + \int\limits_{x-\frac{\Delta}{2}}^{x+\frac{\Delta}{2} - \epsilon}du\;A(u) \int\limits_{u+\epsilon}^{x+\frac{\Delta}{2}}dv\;B(v)\right].\label{leibnitz:sp2}
\end{align}
Then, it is easy to show that:
\begin{align}
	\partial_{x} \left[A(x) \circ B(x) \right] = \lim_{\frac{\Delta}{2} \to \epsilon} \frac{1}{(\Delta - \epsilon)^{2}} &\left[A(x+\nicefrac{\Delta}{2})\int\limits_{x-\frac{\Delta}{2}}^{x+\frac{\Delta}{2} -\epsilon} dv \; B(v) - A(x-\nicefrac{\Delta}{2})\int\limits_{x-\frac{\Delta}{2} +\epsilon}^{x+\frac{\Delta}{2}} dv \; B(v)\right. \nonumber\\
	&\left. + \int\limits_{x-\frac{\Delta}{2}}^{x+\frac{\Delta}{2} -\epsilon} du \; A(u) B(x+\nicefrac{\Delta}{2}) - \int\limits_{x-\frac{\Delta}{2} +\epsilon}^{x+\frac{\Delta}{2}} du \; A(u) B(x-\nicefrac{\Delta}{2})\right]. \label{leibnitz:lhs}
\end{align}
Next, we find for the first term in the right hand side of \eqref{leibnitz:rule}:
\begin{align}
	\partial_{x} \left[A(x) \right] \circ B(x) &= \lim_{\frac{\Delta}{2} \to \epsilon} \frac{1}{(\Delta - \epsilon)^{2}}\left[\int\limits_{x-\frac{\Delta}{2}+\epsilon}^{x+\frac{\Delta}{2}}du\; \partial_{u}A(u) \int\limits_{x-\frac{\Delta}{2}}^{u-\epsilon}dv \;B(v) + \int\limits_{x-\frac{\Delta}{2}}^{x+\frac{\Delta}{2} - \epsilon}du \; \partial_{u}A(u) \int\limits_{u+\epsilon}^{x+\frac{\Delta}{2}}dv\;B(v)\right]\nonumber\\
&=\lim_{\frac{\Delta}{2} \to \epsilon} \frac{1}{(\Delta - \epsilon)^{2}} \left[A(x+\nicefrac{\Delta}{2})\int\limits_{x-\frac{\Delta}{2}}^{x+\frac{\Delta}{2} -\epsilon} dv \; B(v) - A(x-\nicefrac{\Delta}{2})\int\limits_{x-\frac{\Delta}{2} +\epsilon}^{x+\frac{\Delta}{2}} dv \; B(v)\right. \nonumber\\
	&\left. - \int\limits_{x-\frac{\Delta}{2} + \epsilon}^{x+\frac{\Delta}{2}} du \; A(u) B(u- \epsilon) + \int\limits_{x-\frac{\Delta}{2}}^{x+\frac{\Delta}{2} - \epsilon} du \; A(u) B(u+ \epsilon)\right]. \label{leibnitz:rhs1}
\end{align}
Similarly, for the second term in the right hand side of \eqref{leibnitz:rule}, we find:
\begin{align}
	A(x) \circ \partial_{x} \left[ B(x) \right] &= \lim_{\frac{\Delta}{2} \to \epsilon} \frac{1}{(\Delta - \epsilon)^{2}}\left[\int\limits_{x-\frac{\Delta}{2}+\epsilon}^{x+\frac{\Delta}{2}}du\; A(u) \int\limits_{x-\frac{\Delta}{2}}^{u-\epsilon}dv \; \partial_{v}B(v) + \int\limits_{x-\frac{\Delta}{2}}^{x+\frac{\Delta}{2} - \epsilon}du \; A(u) \int\limits_{u+\epsilon}^{x+\frac{\Delta}{2}}dv\; \partial_{v}B(v)\right]\nonumber\\
&=\lim_{\frac{\Delta}{2} \to \epsilon} \frac{1}{(\Delta - \epsilon)^{2}} \left[-\int\limits_{x-\frac{\Delta}{2} +\epsilon}^{x+\frac{\Delta}{2}} du A(u) \; B(x- \nicefrac{\Delta}{2}) + \int\limits_{x-\frac{\Delta}{2}}^{x+\frac{\Delta}{2} -\epsilon} du A(u)\; B(x +\nicefrac{\Delta}{2})\right. \nonumber\\
	&\left. + \int\limits_{x-\frac{\Delta}{2} + \epsilon}^{x+\frac{\Delta}{2}} du \; A(u) B(u- \epsilon) - \int\limits_{x-\frac{\Delta}{2}}^{x+\frac{\Delta}{2} - \epsilon} du \; A(u) B(u+ \epsilon)\right]. \label{leibnitz:rhs2}
\end{align}
Finally, summing the terms in \eqref{leibnitz:rhs1} and \eqref{leibnitz:rhs2} we obtain  \eqref{leibnitz:lhs}, thus verifying the Lebnitz rule \eqref{leibnitz:rule}.

\section{Jacobi identity}\label{app:ji}

In this appendix, we consider the Jacobi identity for the quantum algebra \eqref{qa:equal_interval} and \eqref{qa:adjacent_interval}, and address its relation to the consistency conditions for the classical algebra \eqref{sp:Maillet_classical_algebra} derived in \cite{Freidel:1991jx,Freidel:1991jv}.

The Jacobi identity:
\begin{align}
&\left[ T_{1}(u,u';\lambda), \left[ T_{2}(v,v';\mu), T_{3}(w,w';\rho) \right] \right]	+ \mathbb{P}_{13}\mathbb{P}_{23} \left[T_{1}(w,w';\rho), \left[ T_{2}(u,u';\lambda), T_{3}(v,v';\mu) \right] \right] \mathbb{P}_{23}\mathbb{P}_{13}  \nonumber \\
&+ \mathbb{P}_{13}\mathbb{P}_{12} \left[T_{1}(v,v';\mu), \left[ T_{2}(w,w';\rho), T_{3}(u,u';\lambda) \right] \right] \mathbb{P}_{12}\mathbb{P}_{13} =0,\label{sp:JI_all_diff_points}
\end{align}
with $\mathbb{P}$ denoting the permutation operator acting on the auxiliary spaces, is clearly well-defined with respect to the usual operator product for the case where all points $(u,u',v,v',w,w')$ are different. On the other hand, if some of the points $(u,u',v,v',w,w')$ coincide, it involves products of operators at the same point, and is, therefore, a singular expression. Hence, to formulate a well-defined Jacobi identity for the case of possibly coinciding points, one needs to regularize such products.

To this end, we can use Sklyanin's product \eqref{sp:sklyanin_prod_general} and the following property:
\begin{align}
\left[ A_{1}(x) \stackrel{\circ}{,} \left[ A_{2}(x) \stackrel{\circ}{,} A_{3}(x) \right] \right] =  \lim_{\nicefrac{\Delta}{2} \to \epsilon} \frac{1}{\Delta V} \sum\limits_{i=1}^{3!}\int\limits_{\Delta V_{i}} d\zeta_{1} d\zeta_{2} d\zeta_{3} \;  \left[ A_{1}(\zeta_{1}), \left[  A_{2}(\zeta_{2}), A_{3}(\zeta_{3})\right]\right], \label{sp:property3}
\end{align}
to extend the well-defined expression \eqref{sp:JI_all_diff_points} to the case where some arbitrary subset of points may coincide. Thus, we obtain a formula valid for an arbitrary set of points $(x_{1},y_{1},x_{2},y_{2},x_{3},y_{3})$:
\begin{align}
&\left[ T_{1}(x_{1},y_{1};\lambda)\stackrel{\circ}{,}  \left[ T_{2}(x_{2},y_{2};\mu) \stackrel{\circ}{,}  T_{3}(x_{3},y_{3};\rho) \right] \right] \notag \\
&+ \mathbb{P}_{13}\mathbb{P}_{23} \left[T_{1}(x_{3},y_{3};\rho)\stackrel{\circ}{,} \left[ T_{2}(x_{1},y_{1};\lambda)\stackrel{\circ}{,}  T_{3}(x_{2},y_{2};\mu) \right] \right] \mathbb{P}_{23}\mathbb{P}_{13}  \notag \\
&+  \mathbb{P}_{13}\mathbb{P}_{12} \left[T_{1}(x_{2},y_{2};\mu)\stackrel{\circ}{,}  \left[ T_{2}(x_{3},y_{3};\rho)\stackrel{\circ}{,}  T_{3}(x_{1},y_{1};\lambda) \right] \right] \mathbb{P}_{12}\mathbb{P}_{13} = 0.\label{sp:JI_any_points_SP}
\end{align}
It is clear that when all points $(x_{1},y_{1},x_{2},y_{2},x_{3},y_{3})$ are different, the expression \eqref{sp:JI_any_points_SP} trivially reduces to \eqref{sp:JI_all_diff_points}. We emphasize that, although not explicitly indicated, the formula \eqref{sp:JI_any_points_SP} depends on the regularization parameter $\epsilon$ (see the definitions  \eqref{sp:sklyanin_prod} and \eqref{sp:sklyanin_prod_general}) and, therefore, is a well-defined expression.

The general conditions imposed by the Jacobi identity \eqref{sp:JI_any_points_SP} on the quantum algebra of transition matrices \eqref{qa:equal_interval} and \eqref{qa:adjacent_interval} are not very enlightening. However, as we show in the following, they lead in the classical limit to the same consistency conditions obtained in \cite{Freidel:1991jx,Freidel:1991jv}. For simplicity, we restrict the analysis to the simpler case \cite{Freidel:1991jv} involving only bosonic fields, and for which the matrices $a_{12}$, $d_{12}$ and $b_{12}=c_{21}$ encoding the quantum algebra \eqref{qa:equal_interval} and \eqref{qa:adjacent_interval} depend only on the spectral parameters. In this case, using the quantum algebra \eqref{qa:equal_interval} and \eqref{qa:adjacent_interval} to evaluate the Jacobi identity \eqref{sp:JI_any_points_SP} for all possible combinations of intervals, i.e., equal, adjacent and mixed, we can easily derive in the classical limit the following Yang-Baxter-like constraints:
\begin{align}
	\left[ a_{12}(\lambda,\mu), a_{13}(\lambda,\nu) \right] + \left[ a_{12}(\lambda,\mu), a_{23}(\mu,\nu) \right] + \left[ a_{13}(\lambda,\nu), a_{23}(\mu,\nu) \right] &= 0,\label{sp:YB_a} \\
\left[ d_{12}(\lambda,\mu), d_{13}(\lambda,\nu) \right] + \left[ d_{12}(\lambda,\mu), d_{23}(\mu,\nu) \right] + \left[ d_{13}(\lambda,\nu), d_{23}(\mu,\nu) \right] &= 0,\label{sp:YB_d}\\
\left[ b_{12}(\lambda,\mu), d_{13}(\lambda,\nu)\right] + \left[b_{32}(\nu,\mu), d_{13}(\lambda,\nu)\right] + \left[ b_{32}(\nu,\mu),b_{12}(\lambda,\mu) \right] &=0,\label{sp:YB_bd}\\
\left[ a_{32}(\nu,\mu), c_{21}(\mu,\lambda)\right] + \left[a_{32}(\nu,\mu), c_{31}(\nu,\lambda)\right] + \left[ c_{31}(\nu,\lambda),c_{21}(\mu,\lambda) \right] &=0.\label{sp:YB_ac}
\end{align}
Moreover, we also obtain the classical relation $b_{12}(\lambda,\mu) = c_{21}(\mu,\lambda)$ and the antisymmetry of the parameters $a_{12}(\lambda,\mu)$ and $d_{12}(\lambda,\mu)$ in the description of the classical algebra under the permutation of the auxiliary spaces corresponding to the spectral parameters $\lambda$ and $\mu$.

Finally, we note that two properties enjoyed by Sklyanin's product render the aforementioned calculations a mere repetition of the computation originally performed in \cite{Freidel:1991jv}. Namely, the commutators of operator-valued functions endowed with Sklyanin's product \eqref{sp:sklyanin_prod_general} satisfy the following standard relations:
\begin{align}
	\left[ A_1(x) \stackrel{\circ}{,} A_2(x) \circ A_3(x) \right] &= A_2(x) \circ \left[ A_1(x) \stackrel{\circ}{,}  A_3(x) \right] + \left[ A_1(x) \stackrel{\circ}{,} A_2(x) \right] \circ A_3(x),\label{sp:property1} \\
\left[ A_1(x) \stackrel{\circ}{,}\left[ A_2(x) \stackrel{\circ}{,} A_3(x) \right] \right] &= \left[ A_1(x) \stackrel{\circ}{,} \alpha A_2(x) \circ A_3(x) + A_3(x) \circ A_2(x)  \beta\right],\label{sp:property2}
\end{align}
where \eqref{sp:property2} holds provided the commutator of two operator-valued functions is of the form:
\begin{align*}
\left[ A_2(x) \stackrel{\circ}{,} A_3(x) \right] = \alpha A_2(x) \circ A_3(x) + A_3(x) \circ A_2(x)  \beta,
\end{align*}
with $\alpha, \beta $ being arbitrary constants. Therefore, we omit tedious computational details.

\phantomsection
\addcontentsline{toc}{section}{References}
\bibliographystyle{utphys}
\bibliography{references}

\end{document}